\documentclass[twocolumn,preprintnumbers,prd,tightenlines,nofootinbib]{revtex4}
\usepackage{amsmath,bm}
\usepackage{graphicx}
\def\mi{{\mathrm i}}

\newbox\charbox
\newbox\slabox
\def\s#1{{      
        \setbox\charbox=\hbox{$#1$}
        \setbox\slabox=\hbox{$/$}
        \dimen\charbox=\ht\slabox
        \advance\dimen\charbox by -\dp\slabox
        \advance\dimen\charbox by -\ht\charbox
        \advance\dimen\charbox by \dp\charbox
        \divide\dimen\charbox by 2
        \raise-\dimen\charbox\hbox to \wd\charbox{\hss/\hss}
        \llap{$#1$}
}}
\begin{document}

\title{Direct numerical integration of one-loop 
Feynman diagrams for $N$-photon amplitudes}

\author{Wei Gong}
\affiliation{
Institute of Theoretical Science,
University of Oregon,
Eugene, OR  97403-5203, USA
}
\author{Zolt\'an Nagy}
\affiliation{
Terascale Physics Analysis Center,
DESY,
Notkestrasse 85,
22607 Hamburg, Germany
}
\author{Davison E.\ Soper}
\affiliation{
Institute of Theoretical Science,
University of Oregon,
Eugene, OR  97403-5203, USA{}
}

\preprint{DESY 08-196}

\begin{abstract}
One approach to the calculation of cross sections for infrared-safe observables in high energy collisions at next-to-leading order is to perform all of the integrations, including the virtual loop integration, by Monte Carlo numerical integration. In a previous paper, two of us have shown how one can perform such a virtual loop integration numerically after first introducing a Feynman parameter representation. In this paper, we perform the integration directly, without introducing Feynman parameters, after suitably deforming the integration contour. Our example is the $N$-photon scattering amplitude with a massless electron loop. We report results for $N = 6$ and $N = 8$.
\end{abstract}

\date{17 December 2008}

\maketitle

\section{Introduction}
\label{sec:intro}

The calculation of cross sections in the Standard Model and its extensions at next-to-leading order in perturbation theory inevitably involves computing virtual loop Feynman diagrams. In a process in which 2 partons scatter to produce $n$ partons we integrate over the momenta $\{p_1, \dots, p_n\}$ of the $n$ final state partons. In a leading order calculation, for each choice of $\{p_1, \dots, p_n\}$, we multiply the desired measurement function for that point by the squared tree level matrix element evaluated at that point. At next-to-leading order, we need also the one loop matrix element times the complex conjugate of the tree level matrix element, plus the complex conjugate of this product. The standard method for this kind of calculation involves computing the one loop matrix element as a whole: for the chosen $\{p_1, \dots, p_n\}$, we compute the integral that represents the loop graphs, with infrared and ultraviolet subtractions as necessary. The method for calculating this integral typically involves representing the integral in terms of ``master integrals,'' whose values are known. There has recently been very significant progress in developing this method \cite{Ossolaetal, Ellisetal, Bergeretal}.

There is another possibility. The loop graph is an integral over a loop momentum $l$. One can write the integration over $\{p_1, \dots, p_n\}$ and $l$ as a single integration, so that for each choice of $\{p_1, \dots, p_n\}$ in a Monte Carlo style integration, we also choose a momentum $l$. Then we multiply the measurement function and the complex conjugate tree amplitude by the {\em integrand} of the loop amplitude evaluated at $(\{p_1, \dots, p_n\},l)$. The loop amplitude contains singular factors $1/((l - Q_i)^2 + i\epsilon)$. These singularities can be partly avoided by deforming the integration contour into the complex $l$ space. Thus one needs to specify what the deformed integration contour is to be.

There are some possible variations on this method. In one variation, the integral over the energy $l^0$ is performed analytically by closing the integration contour, leaving a three dimensional integration over $\vec l$. In another variation, the loop integral is re-expressed using Feynman parameters $x$, so that we have an integration over either $l$ and $x$ or just $x$. Then we integrate over a complex contour in $x$. In any of these variations, the loop integral is evaluated by numerical Monte Carlo integration along with the integration over $\{p_1, \dots, p_n\}$, so we may refer to this as the numerical Monte Carlo method.

The numerical Monte Carlo method was implemented in Ref.~\cite{beowulf} for three jet observables in electron-positron annihilation. Here the loop integral is expressed as a three dimensional integration over $\vec l$. The variant of the numerical Monte Carlo method in which the loop integral is expressed as an integration over Feynman parameters $x$ has been successfully implemented in NLO calculations by Lazopoulos, Melnikov and Petriello \cite{Lazopoulos1} for $pp \to ZZZ +X$ and by Lazopoulos, McElmurry, Melnikov and Petriello \cite{Lazopoulos2} for $pp \to t\bar t Z +X$. Furthermore, Anastasiou, Beerli and Daleo have used this method to compute the two loop amplitudes needed for $g g \to h$ mediated by a heavy quark or a scalar quark \cite{Anastasiou}. In these methods the infrared singularities are eliminated by the method of sector decomposition \cite{sectordecompose}. For the contour deformation in $x$ space, these authors follow the prescription given in Ref.~\cite{NSnumerical}.

\begin{figure}
\centerline{
\includegraphics[width = 6 cm]{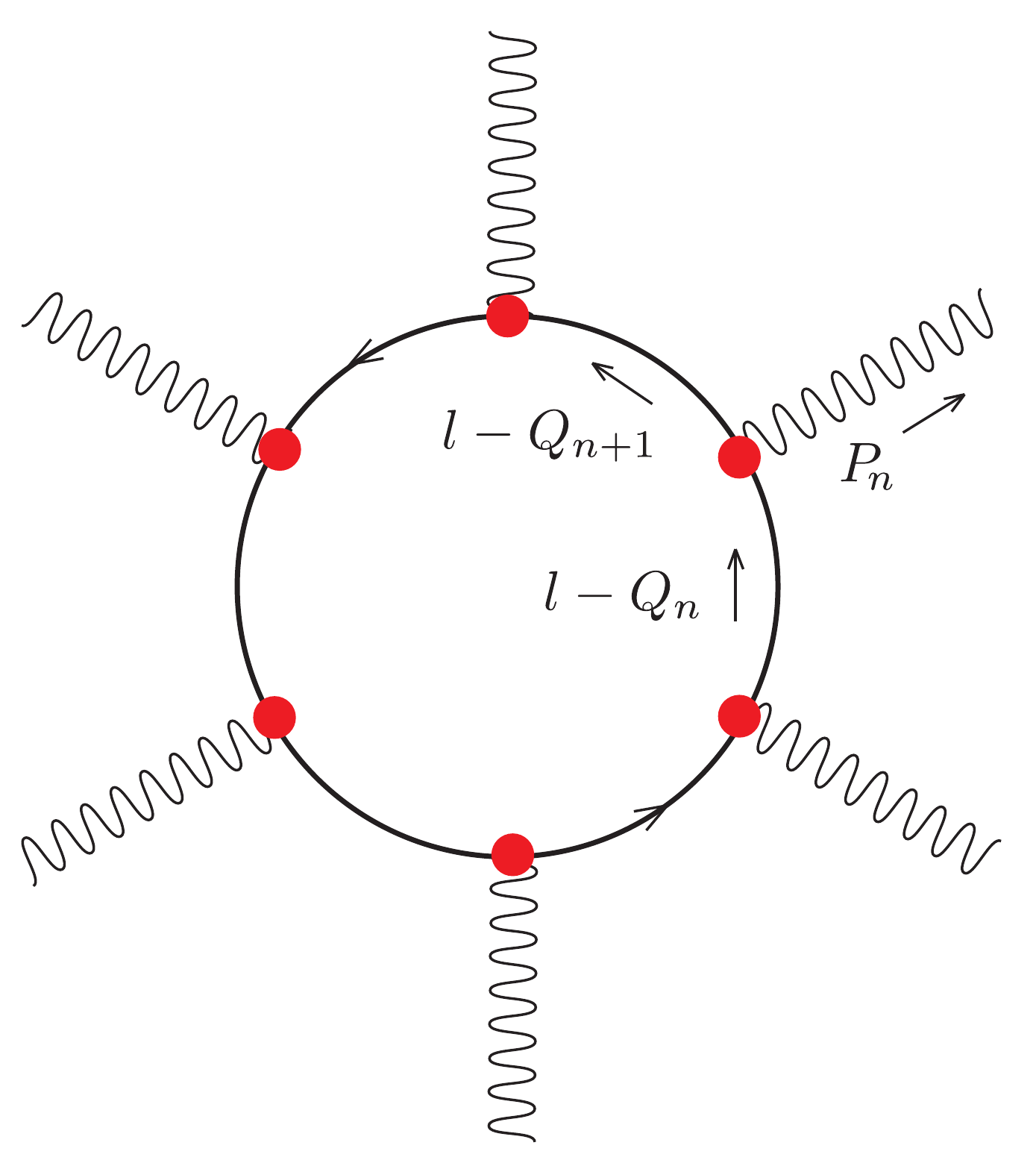}
}
\medskip
\caption{
Feynman diagram for the $N$-photon amplitude.
}
\label{fig:Nphoton}
\end{figure}

In Ref.~\cite{NSnumerical}, two of the present authors explored the numerical Monte Carlo method using the Feynman parameter representation by taking as an example the amplitude for $\gamma + \gamma \to (N-2)\,\gamma$ through a (massless) electron loop, as depicted in Fig.~\ref{fig:Nphoton}. The integral to be evaluate has the form
\begin{equation}
{\cal M} = \int\! \frac{d^4 l}{(2\pi)^4}\ (-\mi e)^N N(l)
\prod_{i=1}^N \frac{\mi}{(l - Q_i)^2 + \mi 0}
\;\;,
\label{eq:calM0}
\end{equation}
where $N(l)$ is the numerator function. This particular amplitude is useful as a test case because the loop integrand has infrared singularities, but (for a generic choice of $\{p_1, \dots, p_n\}$) these singularities are integrable. Thus infrared subtractions are not needed and one can test the integration method without confronting a subtraction method at the same time. The integral Eq.~(\ref{eq:calM0}) was re-expressed using Feynman parameters $x$, giving an integral with the form
\begin{equation}
\begin{split}
{\cal M} ={}& 
- m_0^2 e^N \Gamma(N+1)
\int\! \frac{d^4 \ell}{(2\pi)^4}\
\frac{1}{[1 +  \ell\cdot\ell]^{N+1}}
\\&\times
\int_{C}d x\, 
\left(\sum_{i=1}^N x^i\right)^{\!\!N-3}\
	\frac{N(l(x,\ell))}{\big[\Lambda^{2}(x)\big]^{N-1}}
\;\;,
\label{eq:lxspacedeformed}
\end{split}
\end{equation}
where $\ell$ is a translated and Wick rotated loop momentum, $\ell\cdot\ell$ is the euclidian square of $\ell$, and $\Lambda^{2}(x)$ is a quadratic function of the Feynman parameters $x$. The Feynman parameters are integrated over a certain deformed contour $C$.

Although the numerical Monte Carlo integral (\ref{eq:lxspacedeformed}) is guaranteed to converge as the number of integration points becomes arbitrarily large, it needed to be proved that the method could produce the right answer with a practical number of integration points. For this reason $\cal M$ was evaluated at specified points $\{p_1, \dots, p_n\}$ by numerical Monte Carlo integration. The practicality test showed that one could obtain results for $N=6$. For some helicity choices, the six gluon amplitude was known analytically \cite{Mahlon} and the numerical results agreed with the analytical answer. For other helicity choices, the results have been confirmed by independent analytical calculations by Binoth, Heinrich, Gehrmann and Mastrolia \cite{Binoth} and by Ossola, Papadopoulos and Pittau \cite{Ossola}.

In this paper, we ask whether one could use Monte Carlo numerical integration to perform the integration in Eq.~(\ref{eq:calM0}) directly, without transforming the integral into the Feynman parameter form (\ref{eq:lxspacedeformed}). In a practical application for which infrared subtractions are needed, as in Refs.~\cite{Lazopoulos1, Lazopoulos2, Anastasiou}, one would lose the possibility of constructing the subtractions with the sector decomposition method. On the other hand, it is known \cite{NSsubtractions} how to construct the subtractions directly in momentum space in a fashion that is analogous to that used for real emission diagrams. If one were to perform the integration in Eq.~(\ref{eq:calM0}) directly, there would be a certain advantage of simplicity. Additionally, one would avoid having a denominator $\Lambda(x)$ raised to a high power, which can create numerical convergence difficulties.

The real apparent advantage of the Feynman parameter integration is that $\Lambda(x)$ is simply a quadratic function of $x$, so that it is easy to find a contour deformation that keeps us away from the zeros of the denominator. With Eq.~(\ref{eq:calM0}), the denominator is a product of factors, each of which vanishes on a different surface in $l$ space. For this reason, it is not immediately evident how to deform the integration contour in $l$. In the subsequent sections, we lay out a contour deformation with the required properties. We then apply the same practicality test that was used for the form (\ref{eq:lxspacedeformed}). We find that the direct form (\ref{eq:calM0}) gives us somewhat better numerical convergence than does the Feynman parameter form (\ref{eq:lxspacedeformed}). We do not know if the direct form is better in real applications. Indeed, it may well prove more useful to use the style of calculation in which the integral $\cal M$ is evaluated as a whole in terms of master integrals. Alternatively, it may be that one method is better for some applications while other methods work better in other applications. We offer a method for performing the integral (\ref{eq:calM0}) directly in this paper in order to extend the range of available choices.

\section{Integrating on a deformed contour}

The integrand in Eq.~(\ref{eq:calM0}) has singularities on the surfaces $(l - Q_i)^2 = 0$. In order to avoid these singularities, we can deform the integration contour so that the loop momentum has an imaginary part. Call the complex loop momentum $\ell$ and let
\begin{equation}
\ell^\mu(l) = l^\mu + \mi \kappa^\mu(l)
\;\;.
\end{equation}
Here $l^\mu$ and $\kappa^\mu$ are the real and imaginary parts of $\ell^\mu$ and $\kappa$ is a function of $l$. With this notation, the integral is
\begin{equation}
{\cal M} = \int\! \frac{d^4 l}{(2\pi)^4}\ (-\mi e)^N 
\det(\partial \ell/\partial l)\
N(\ell(l))
\prod_{i=1}^N \frac{\mi}{(\ell - Q_i)^2}
\;\;.
\label{eq:calM1}
\end{equation}
In moving the integration contour we make use of the multidimensional version of the widely used one dimensional contour integration formula. A simple proof is given in the second paper of Ref.~\cite{beowulf}. The essence of the theorem is that we can move the integration contour as long as we start in the direction indicated by the $+\mi 0$ prescription and do not encounter any singularities of the integrand along the way.

In order to see what is required for the deformation, we consider a family of deformations, specified by
\begin{equation}
\kappa^\mu(l) = \lambda(l) \, \kappa^\mu_0(l)
\;\;.
\end{equation}
We imagine starting with an infinitesimal $\lambda$ and then increasing it to its final value, $\lambda_{\rm f}(l)$. Thus we consider $0 < \lambda(l) < \lambda_{\rm f}(l)$. The denominator corresponding to propagator $i$ is
\begin{equation}
\begin{split}
(l - Q_i + \mi \kappa(l))^2 ={}& 
(l - Q_i)^2 - \lambda(l)^2\,\kappa_0(l)^2 
\\&
+ 2 \mi\,\lambda(l)\, (l - Q_i)\cdot \kappa_0(l)
\;\;.
\label{eq:denomform}
\end{split}
\end{equation}
Our first requirement is that we start the deformation in the direction specified by the $+\mi 0$ prescription. This means that on any of the of the surfaces $(l - Q_i)^2 = 0$ we have $(l - Q_i)\cdot \kappa_0(l) \ge 0$.

Consider a a point $l$ on the cone $(l - Q_i)^2 = 0$. The condition $(l - Q_i)\cdot \kappa_0(l) > 0$ has a simple geometrical interpretation: that $\kappa_0(l)$ points toward the interior of the cone. To see this, we consider  the point $l + \lambda \kappa_0$, where  $\lambda$ is infinitesimal and positive. Points on the interior of the cone have $(l + \lambda \kappa_0 - Q_i)^2 > 0$. Expanding to first order in $\lambda$ and using $(l - Q_i)^2 = 0$, we have $2\lambda \kappa_0\cdot (l - Q_i) > 0$. Similarly, the condition $(l - Q_i)\cdot \kappa_0(l) = 0$ has the geometrical interpretation that $\kappa_0(l)$ is tangent to the cone.

We want to escape from the singularities if we can. This means that on $(l - Q_i)^2 = 0$ we would like $(l - Q_i)\cdot \kappa_0(l) > 0$. This is easy as long as there is only one cone involved. We simply need to find a vector $\kappa_0$ that points toward the interior of the cone. No deformation is possible for the point $l = Q_i$. We cannot have $(l - Q_i)\cdot \kappa_0(l) > 0$ if $(l - Q_i) = 0$. Thus the point $l = Q_i$ is a pinch singularity, meaning that we cannot deform the contour to get away from it.

At the intersection of two cones, deforming the contour is a little more subtle. Consider two cones $(l - Q_i)^2 = 0$ and $(l - Q_j)^2 = 0$ and suppose that $K \equiv Q_j - Q_i$ is a timelike vector. On the intersection of these two cones, we need a vector $\kappa_0(l)$ that points towards the interior of both cones. It is geometrically evident that this is possible. If $K$ is a spacelike vector, we also need a vector $\kappa_0(l)$ that points towards the interior of both cones. It is also geometrically evident that this is possible. 

If $K \equiv Q_j - Q_i$ is a lightlike vector, the cones meet along a line $l - Q_i = x K$. If $x > 1$ or $x < 0$, the inside of one of the cones is inside of the other, so that there is a range of vectors $\kappa_0(l)$ that point toward the interior of both cones. However if $0<x<1$, the inside of one cone is outside of the other, so that there is no vector $\kappa_0(l)$ that points toward the interior of both cones. Specifically, for $l = Q_i + x K = Q_j - (1-x) K$ with $0 < x < 1$, we need $x K \cdot \kappa_0(l) \ge 0$ and $(1-x)K \cdot \kappa_0(l) \le 0$. The best that we can do is have  $x K \cdot \kappa_0(l) = (1-x) K \cdot \kappa_0(l) = 0$. Since $K^2 = 0$, this is possible with $\kappa_0(l) = c(x) K$. Thus the contour is pinched: we can deform along $K$, but not in any other direction. With this deformation, we do not escape from the singularity. This pinch singularity is called the collinear singularity.

\section{Preview of the deformation}

\begin{figure}
\centerline{
\includegraphics[width = 6 cm]{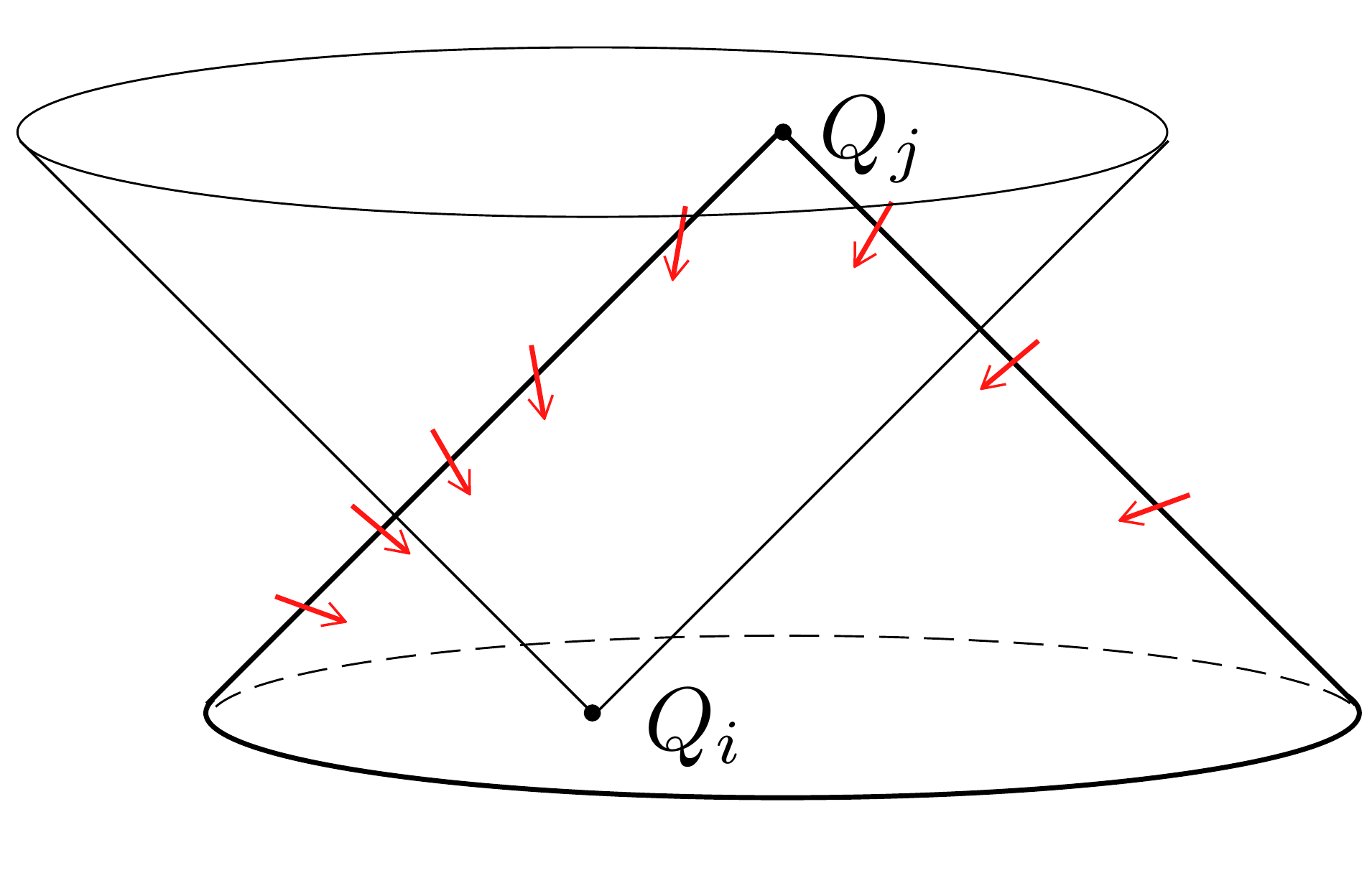}
}
\medskip
\caption{
Direction of the deformation $\kappa_0 = -  c\, (l-Q_i)$ for selected points on the backward light cone from $Q_j$ when $Q_j - Q_i$ is a timelike vector with a positive time component. The arrows, which represent the direction of $\kappa_0$, point to the interior of the backward light cone from $Q_j$, so that $\kappa_0 \cdot (l-Q_j) > 0$.
}
\label{fig:cones}
\end{figure}

In the subsequent sections, we define a contour deformation quite precisely. In this section, we provide an informal statement of the main idea. Consider two cones $(l - Q_i)^2 = 0$ and $(l - Q_j)^2 = 0$ and suppose that $K \equiv Q_j - Q_i$ is a timelike vector with $K^0 > 0$. Let
\begin{equation}
\label{eq:idea}
\kappa_0 = -  c\, (l-Q_i)
\;\;.
\end{equation}
The coefficient $c$ is a non-negative function of $l$. We want
\begin{equation}
\label{eq:requirementi}
\kappa_0\cdot (l-Q_i) \ge 0
\end{equation}
on the surface $(l-Q_i)^2 = 0$. But
\begin{equation}
\kappa_0\cdot (l-Q_i) = -c\, (l-Q_i)^2
\;\;,
\end{equation}
so this requirement is automatically met. We also want
\begin{equation}
\label{eq:requirementj}
\kappa_0\cdot (l-Q_j) \ge 0
\end{equation}
on the surface $(l-Q_j)^2 = 0$. There are two cases to consider. For the backward light cone from $Q_j$, we simply note that the point $Q_i$ lies inside this backward light cone. Thus for $l$ on the backward light cone from $Q_j$, $\kappa_0$ points to the interior of this light cone and $\kappa_0\cdot (l-Q_j) > 0$. This is illustrated in Fig.~\ref{fig:cones}. For the forward light cone from $Q_j$, $\kappa_0$ points in the wrong direction if $c > 0$. However, we can have $\kappa_0\cdot (l-Q_j) = 0$ by taking $c = 0$ on the forward light cone from $Q_j$.

\begin{figure}
\centerline{
\includegraphics[width = 8 cm]{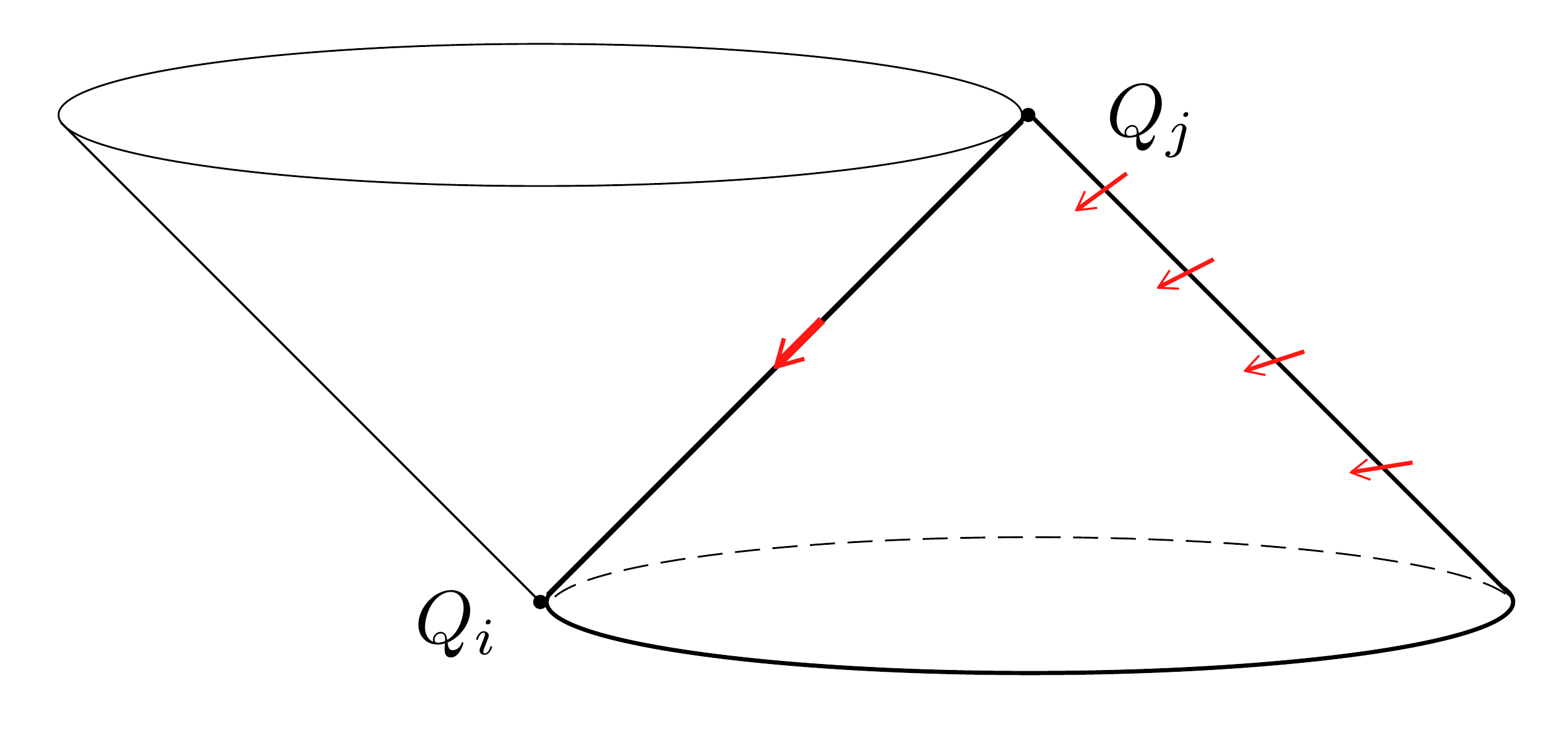}
}
\medskip
\caption{
Direction of the deformation $\kappa_0 = -  c\, (l-Q_i)$ for selected points on the backward light cone from $Q_j$ when $Q_j - Q_i$ is a lightlike vector with a positive time component. For a generic point on the backward light cone from $Q_j$, the arrows, which represent the direction of $\kappa_0$, point to the interior of the light cone, so that $\kappa_0 \cdot (l-Q_j) > 0$. On the lightlike line along the intersection of the two cones, $\kappa_0$ is parallel to $(l-Q_j)$, so that $\kappa_0 \cdot (l-Q_j) = 0$.
}
\label{fig:cones2}
\end{figure}

This construction also works if $K$ is lightlike with $K^0 > 0$. The only difference is that now the point $l = Q_i$ lies {\em on} the backward light cone from $Q_j$. This means that $\kappa_0\cdot (l-Q_j) > 0$ on the backward light cone from $Q_j$ except along the line from $Q_i$ to $Q_j$, where $\kappa_0\cdot (l-Q_j) = 0$. That is, we escape from the singularity everywhere except along the collinear pinch singular line. This is illustrated in Fig.~\ref{fig:cones2}.

We see that the contour deformation of Eq.~(\ref{eq:idea}) achieves  $\kappa_0\cdot (l-Q_j) \ge 0$ on $(l-Q_j)^2 = 0$ and $\kappa_0\cdot (l-Q_i) \ge 0$ on $(l-Q_i)^2 = 0$. What we need is to replace the $\ge$ signs by $>$ signs on the light cones except where the contour is pinched. We can achieve this by including several terms of the form suggested by Eq.~(\ref{eq:idea}) plus, in certain regions of $l$, some terms pointing in a fixed timelike direction. We specify this choice in subsequent sections.

\section{Defining the integral}

As in Ref.~\cite{NSnumerical}, we wish to compute the amplitude in quantum electrodynamics for scattering of two photons to produce $N-2$ photons by means of a (massless) electron loop. This process is illustrated in Fig.~\ref{fig:Nphoton}. The integral is given in Eq.~(\ref{eq:calM0}). Electron line $n$ in the loop carries momentum $l - Q_n$, where $Q_n$ is fixed and we integrate over $l$. The momentum carried out of the graph by external photon $n$ is 
\begin{equation}
P_n = Q_{n+1} - Q_n \;\;,
\label{Qndef}
\end{equation}
with $P_n^2 = 0$. 

We attack this simple problem because infrared counter terms are not needed: the original integral is infrared finite. The propagator denominators provide factors that would lead to logarithmic divergences after integration over the soft and collinear regions. However, these divergences are cancelled. For each electron line there is a factor $(\s l - \s Q_n)$. Thus the numerator provides a factor that removes the soft divergence from the integration region $(l - Q_n)\to 0$. Similarly at each vertex there is a factor $(\s l - \s Q_{n+1})\ \s\epsilon_n(P_n)\ (\s l - \s Q_{n})$, where $\epsilon_n(P_n)$ is the polarization vector of the photon. In the collinear limit $(l - Q_n) \to x P$, this gives a factor  $-x(1-x) \s P_n \s\epsilon_n(P_n) \s P_n  = - 2x(1-x) \s P_n\ \epsilon_n(P_n)\cdot P_n$. This vanishes because $\epsilon_n(P_n)\cdot P_n = 0$. Thus the numerator also provides a factor that removes each collinear divergence. The loop integral is also finite in the ultraviolet as long as $N>4$. (For $N=4$ the integral is divergent by power counting, so an ultraviolet counter term is needed.)  

It will prove helpful to adopt a bit more notation.  Two of the momenta $P_i$ are the negatives of the momenta of the two incoming partons. We choose our labels so that these are $P_N$ and $P_A$ (for some $A \ne N$). We define $P = - P_N$ and $\bar P = - P_A$:
\begin{equation}
\begin{split}
P ={}& Q_N - Q_1
\;\;,
\\
\bar P ={}& Q_A - Q_{A+1}
\;\;.
\end{split}
\end{equation}
We choose our reference frame so that the transverse momenta $P_T$ and $\bar P_T$ of the two incoming particles vanish.

\begin{figure}
\centerline{
\includegraphics[width = 8 cm]{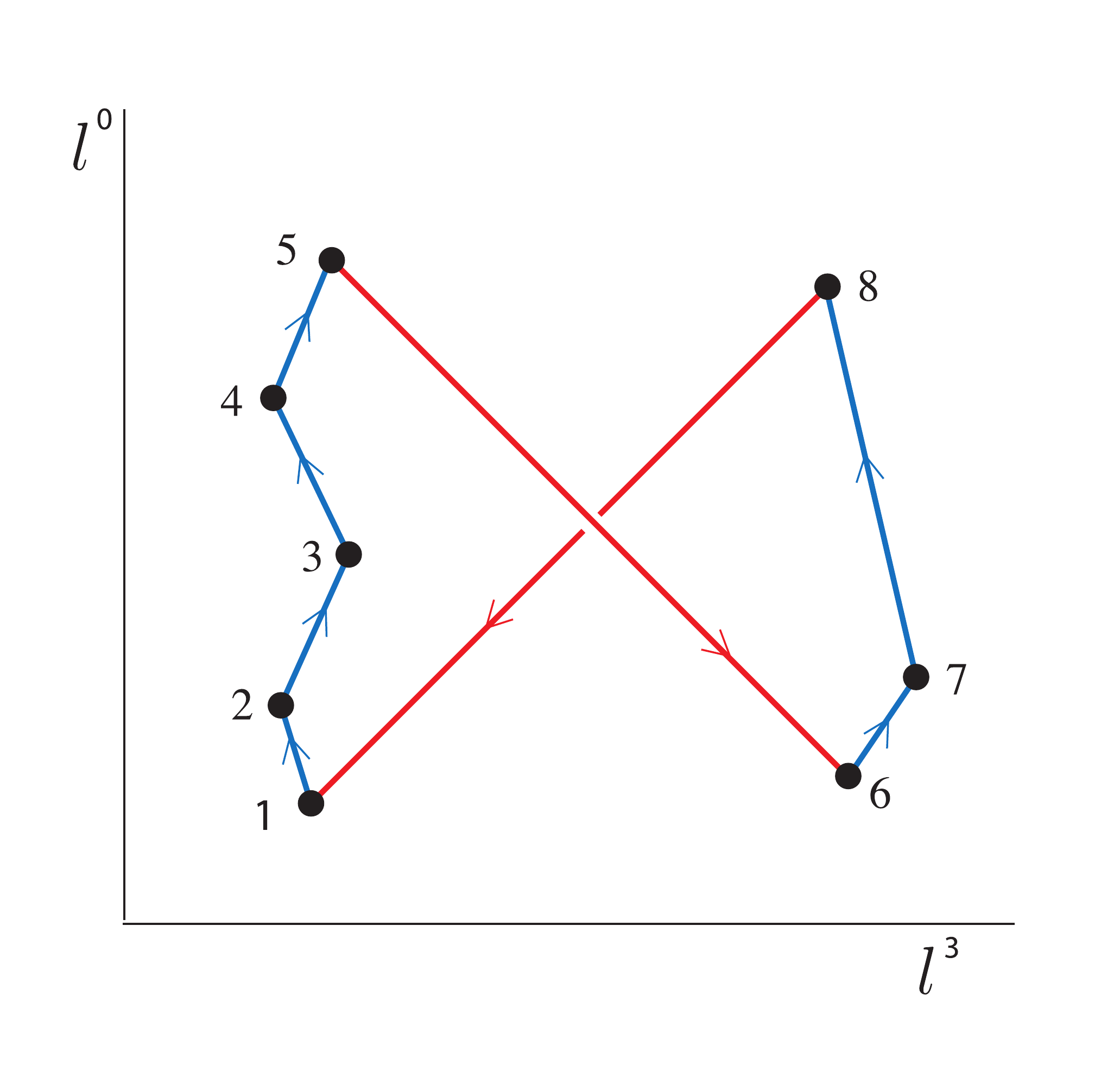}
}
\medskip
\caption{
Kinematics for the $N$-photon amplitude, illustrated for $N = 8$. The sketch shows the $l^0$ and $l^3$ components of the loop momentum $l$. There are also two transverse components that come out of the plane of the paper and are not seen. The points are possible points $l = Q_i$.  The lines $l - Q_i = x P_i$, where $P_i = Q_{i+1} - Q_i$ and $0 \le x \le 1$, are also shown joining the points. The $P_i$ are lightlike momenta. 
}
\label{fig:Qn}
\end{figure}

There are some facts about the kinematics that can be easily be understood with the aid of Fig.~\ref{fig:Qn}. There we see the 0 and 3 components of the momenta $Q_n$ for a sample event. The momenta $P_i = Q_{i+1} - Q_i$ of the external particles join the points. In this illustration, $P_1$, $P_2$, $P_3$, and $P_4$ represent final state particles. Then $P_5$ is the negative of the momentum of an incoming particle. Next, $P_6$ and $P_7$ are the momenta of final state particles and $P_8$ is the negative of the momentum of the other initial state particle. We can see from the figure without the need of an algebraic proof that $Q_b - Q_a$ is inside the positive light cone if only outgoing particles are attached along the loop in the positive loop direction from line $a$ to line $b$, as is the case, for instance for $Q_5 - Q_2$. We see also that $Q_b - Q_a$ is spacelike if precisely one initial state particle and at least one final state particle lie along the loop moving from $a$ to $b$, as is the case, for instance, for $Q_5 - Q_7$.

One can deform away from the singularities $(l - Q_i)^2 = 0$ except where the contour is pinched along the straight lines $l - Q_i = x P_i$ for $0 \le x \le 1$. These are the lines depicted in Fig.~\ref{fig:Qn}. The endpoints of the line segments, where two line segments meet, are the soft singularities. The interiors of the lines, with $0 < x < 1$, represent the collinear singularities. At the soft singularities, $l = Q_i$, the deformation has to vanish,
\begin{equation}
\kappa_0(Q_i) = 0
\;\;.
\end{equation}
Along one of the collinear lines, the deformation has to be parallel to the line,
\begin{equation}
\kappa_0^\mu(Q_i + x P_i) = c(x) P_i^\mu
\;\;,
\end{equation}
for $0 < x < 1$. Thus we will have $(l - Q_i)\cdot \kappa_0(l) = 0$ along one of the collinear lines. Away from these lines we want the denominators {\em not} to vanish. Thus we require
\begin{equation}
(l - Q_i)\cdot \kappa_0(l) > 0
\label{eq:deformcondition}
\end{equation}
on the surface $(l - Q_i)^2 = 0$ except along the collinear lines in this surface. We also demand that $\kappa_0(l)$ be a continuous function. Then Eq.~(\ref{eq:deformcondition}) also holds in a neighborhood of any point on $(l - Q_i)^2 = 0$ excluding the collinear lines.

\section{Geometric arrangement of the light cones}

\begin{figure}
\centerline{
\includegraphics[width = 8 cm]{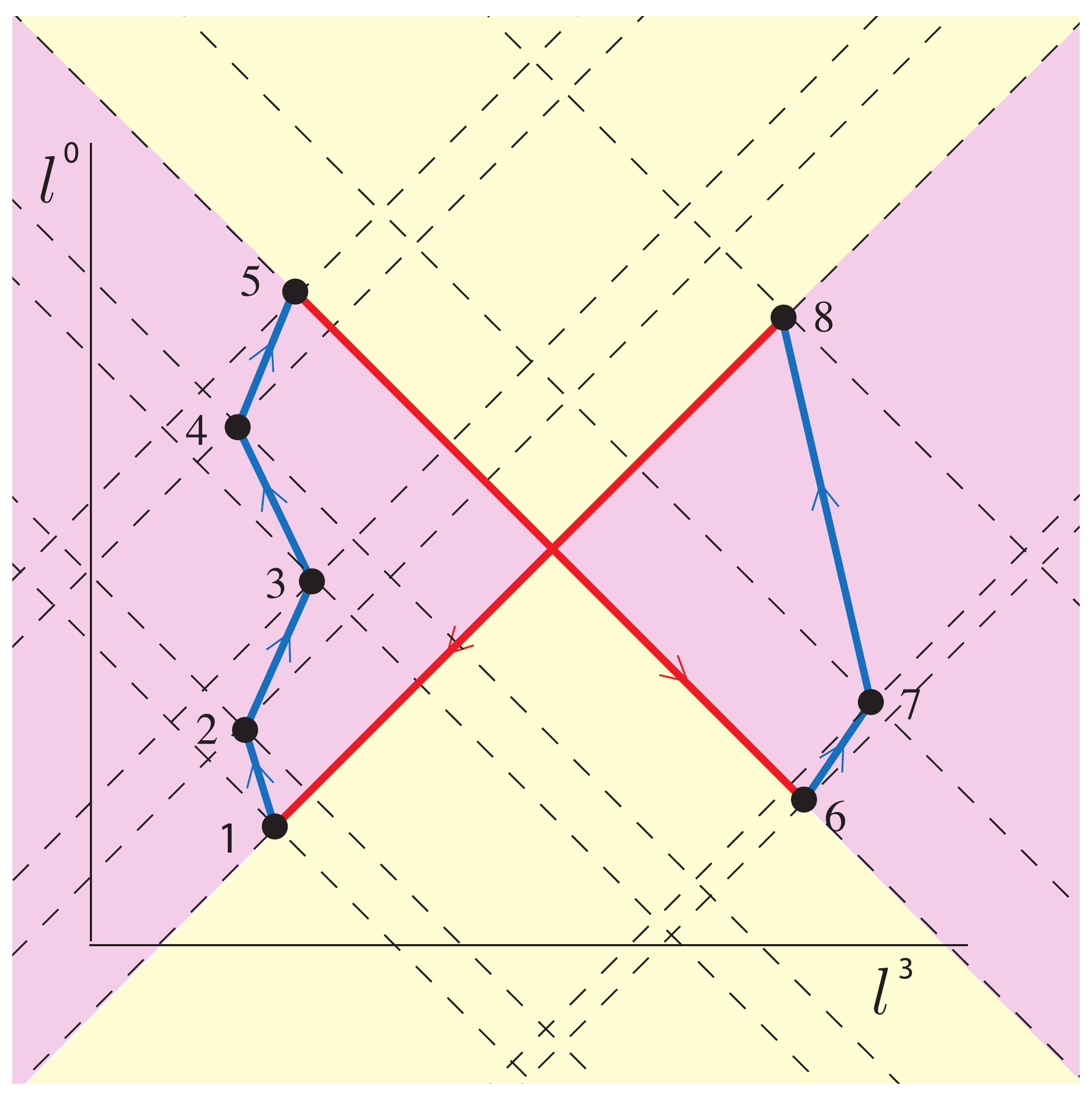}
}
\medskip
\caption{
Kinematics for the $N$-photon amplitude, illustrated for $N = 8$, showing light cones $(l - Q_i)^2 = 0$. In the illustration, $N = 8$ and $A = 5$.
}
\label{fig:Qnplus}
\end{figure}

Fig.~\ref{fig:Qnplus} shows the kinematics as in Fig.~\ref{fig:Qn} but with more information indicated. We show the $l^0$ and $l^3$ components of the loop momentum $l$ with the projections onto the $l^0$-$l^3$ plane of the points $l = Q_n$ indicated. The projections of the light cones $(l - Q_n)^2 = 0$ are the regions between the two dashed lines that pass through each $Q_n$. Four shaded regions are indicated.

Consider light cones with vertices in the left region, those with vertices $\{Q_1, \dots, Q_A\}$. The forward light cone from $Q_i$ is tangent to the backward light cone from $Q_{i+1}$ along the line from $Q_i$ to $Q_{i+1}$. The forward light cone from $Q_i$ intersects the backward light cone from $Q_{j}$ for $j > i + 1$. The forward light cone from $Q_i$ is tangent to the forward light cone from $Q_{i+1}$, while the forward light cones from $Q_{j}$ for $j > i+1$ are nested inside the forward light cone from $Q_i$. The backwards light cones are similarly nested: the backward light cone from $Q_j$ is tangent to the backward light cone from $Q_{j-1}$, while the backwards light cones from $Q_{i}$ for $i < j-1$  are nested inside the backward light cone from $Q_j$.

The light cones with vertices in the right region, with vertices $\{Q_{A+1}, \dots, Q_N\}$, have analogous geometrical relationships with one another.

The forward light cone from a vertex $Q_i$ on the left does not intersect a backward light cone from a vertex $Q_j$ on the right except for $i=1$, $j=N$, for which these cones are tangent along the line from $Q_1$ to $Q_N$. The forward light cone from a $Q_i$ on the right does not intersect a backward light cone from a $Q_j$ on the left except for $i=A+1$, $j=A$, for which these cones are tangent along the line from $Q_A$ to $Q_{A+1}$.

The forward light cones with vertices $Q_1$ and $Q_N$ are tangent, as are forward light cones with vertices $Q_{A+1}$ and $Q_A$. The other forward light cones with vertices in the left region intersect the forward light cones with vertices in the right region. These intersections are in the top region.

The backward light cones with vertices $Q_1$ and $Q_N$ are tangent, as are backward light cones with vertices $Q_{A+1}$ and $Q_A$. The other backward light cones with vertices in the left region intersect the backward light cones with vertices in the right region. These intersections are in the bottom region.

We will make use of these geometrical properties in constructing the deformation. We note that it is possible to have $A = 1$ or $A = N$. In these cases, the picture looks rather different but the properties stated above still hold.

We can define coordinates $x$ and $\bar x$ as follows:
\begin{equation}
\begin{split}
x ={}& \frac{(l - Q_{A+1})\cdot \bar P}{P\cdot \bar P}
\;\;,
\\
\bar x ={}& \frac{(l - Q_{1})\cdot P}{P\cdot \bar P}
\;\;.
\end{split}
\end{equation}
Then the left region is $x < 0$, $\bar x > 0$, the right region is $x > 0$, $\bar x < 0$, the top region is $x > 0$, $\bar x > 0$, and the bottom region is $x < 0$, $\bar x < 0$.

\section{Double parton scattering singularity}
\label{sec:doublesingularity}

\begin{figure}
\centerline{
\includegraphics[width = 8 cm]{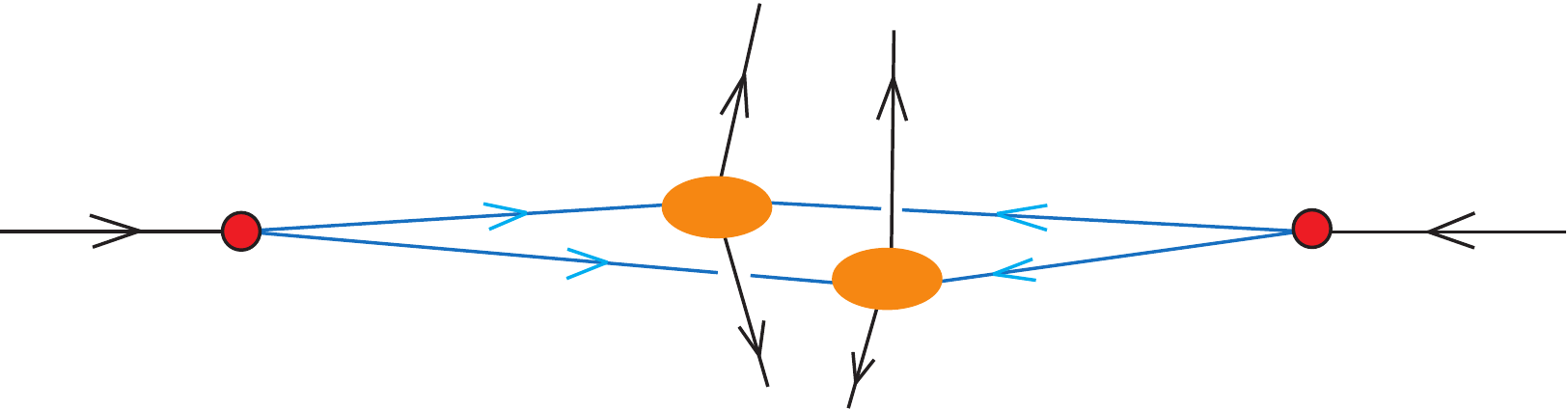}
}
\medskip
\caption{
Illustration of the double parton scattering singularity.
}
\label{fig:dps}
\end{figure}

If the line from $Q_1$ to $Q_N$ intersects the line from $Q_{A+1}$ to $Q_A$, there is a pinch singularity at the intersection that gives a singular integral.\footnote{The amplitude as a function of the external momenta $\{p_1,\dots,p_N\}$ is non-analytic at this point. However, at least for $N = 6$, the amplitude does not become infinite as one approaches the singular point \cite{Bernicot}.} This is the double parton scattering singularity. In general, these lines do not intersect, but they may be close to intersecting.

Physically, the double parton scattering singularity arises when the process illustrated in Fig~\ref{fig:dps} is possible with on-shell intermediate states.  Incoming photon $N$ divides into a collinear electron-positron pair. Incoming photon $A$ divides into a collinear electron-positron pair also. The electron from photon $N$ collides with the positron from photon $A$ so as to produce two or more outgoing photons, which must then have total transverse momentum equal to zero.\footnote{Recall that we choose our reference frame so that the transverse momenta $P$ and $\bar P$ of the two incoming photons vanish.} The electron from photon $A$ collides with the positron from photon $N$ so as to produce a different set of two or more outgoing photons, also with total transverse momentum equal to zero. Thus there is a singularity when the total transverse momentum of some subset of the final state photons, containing at least two photons, has total transverse momentum $p_{\rm T}$ equal to zero. One is close to a double parton scattering singularity when $p_{\rm T}^2 \ll s$.

All points on the line from $Q_{1}$ to $Q_N$ have $\bar x = 0$. The point on this line that also has $x = 0$ is
\begin{equation}
v_{\rm I} = a_{\rm I}\,  Q_1 + (1 - a_{\rm I})\,  Q_N
\;\;,
\end{equation}
where
\begin{equation}
\begin{split}
0 ={}& x\ P \cdot \bar P
\\
={}& [a_I Q_1 + (1-a_I) Q_N - Q_{A+1}]\cdot \bar P
\\
={}& [- a_I P + Q_N - Q_{A+1}]\cdot \bar P
\;\;,
\end{split}
\end{equation}
so
\begin{equation}
\begin{split}
a_{\rm I} = \frac{(Q_N - Q_{A+1})\cdot \bar P}{P\cdot \bar P}
\;\;,
\\
1 - a_{\rm I} = \frac{(Q_{A+1} - Q_1)\cdot \bar P}{P\cdot \bar P}
\;\;.
\end{split}
\end{equation}

Similarly, all points on the line from $Q_{A+1}$ to $Q_A$ have $x = 0$. The point on this line that also has $\bar x = 0$ is
\begin{equation}
v_{\rm II} = a_{\rm II}\, Q_{A+1} + (1 - a_{\rm II})\,  Q_{A}
\;\;,
\end{equation}
where
\begin{equation}
\begin{split}
a_{\rm II} ={}& \frac{(Q_{A} - Q_{1})\cdot P}{P\cdot \bar P}
\;\;,
\\
1 - a_{\rm II} ={}& \frac{(Q_1 - Q_{A+1})\cdot P}{P\cdot \bar P}
\;\;.
\end{split}
\end{equation}

The difference $v_{\rm I} - v_{\rm II}$ is a spacelike vector. The separation between the points is then measured by $- (v_{\rm I} - v_{\rm II})^2$. When  $v_{\rm I} - v_{\rm II}$ is small, the region in which the integrand is almost singular is near
\begin{equation}
\label{eq:vdpsdef}
v \equiv \frac{v_{\rm I} + v_{\rm II}}{2}
\;\;.
\end{equation}

In our computer code, we choose the origin of coordinates for $l$ so that $v = 0$. However, our notation in this paper does not assume this choice.

\section{The deformation}
\label{sec:thedeformation}

In this section, we propose a specific deformation function $\kappa_0(l)$. The final deformation $\kappa(l)$ will be proportional to $\kappa_0(l)$, but with its size adjusted to ensure that it is not too large. We will first simply state the definition of $\kappa_0(l)$. Then we will explain the rational for its various parts. 

\subsection{The general formula}
\label{sec:deformationformula}

The definition is
\begin{equation}
\label{eq:grand}
\kappa_0 = -\sum_{j=1}^N c_j (l-Q_j)
+\tilde c_+\,(P + \bar P)
-\tilde c_-\,(P + \bar P)
\;\;.
\end{equation}
The coefficients $c_j$ and $\tilde c_\pm$ are non-negative functions of $l$. For the generic case $1 < A < N-1$ we define
\begin{equation}
\begin{split}
\label{eq:cjgeneral}
c_j ={}& h_-(l-Q_{j-1})\,h_+(l-Q_{j+1})\,h_-(l-Q_{N})
\\&\times
h_+(l-Q_{A+1})\,
g(l)\,
\quad j \in \{2,\dots,A-1\}\;,
\\
c_j ={}& h_-(l-Q_{j-1})\,h_+(l-Q_{j+1})\,
h_-(l-Q_{A})
\\&\times
h_+(l-Q_{1})\,
g(l)\,
\quad j \in \{A+2,\dots,N-1\}\;,
\\
c_1 ={}& 
h_+(l-Q_{2})\,
h_-(l-Q_{N-1})\,h_+(l-Q_{A+1})\,
g(l)
\;,
\\
c_A ={}& h_-(l-Q_{A-1})\,
h_+(l-Q_{A+2})\,h_-(l-Q_{N})\,
g(l)\, 
\;,
\\
c_{A+1} ={}& h_+(l-Q_{A+2})\,
h_-(l-Q_{A-1})\,h_+(l-Q_{1})\,
g(l)\,
\;,
\\
c_N ={}& h_-(l-Q_{N-1})\,
h_+(l-Q_{2})\,h_-(l-Q_{A})\,
g(l)\, 
\;,
\\
\tilde c_+ ={}& h_-(l-Q_{A})\,h_-(l-Q_{N})
\\&\times
(x+\bar x)\,\theta(x + \bar x > 0)\
g_-(l)
\;,
\\
\tilde c_- ={}& h_+(l-Q_{1})\,h_+(l-Q_{A+1})
\\&\times
[-(x+\bar x)]\,\theta(x + \bar x < 0)\
g_+(l)
\;.
\end{split}
\end{equation}
We define the functions $h_\pm(l)$, $g_\pm(l)$, and $g(l)$ below. For the special case $A=1$, we define
\begin{equation}
\begin{split}
\label{eq:cjnAis1}
c_j ={}& h_-(l-Q_{j-1})\,h_+(l-Q_{j+1})\,
h_-(l-Q_{1})
\\&\times
h_+(l-Q_{1})\,
g(l)\,
\quad j \in \{3,\dots,N-1\}\;,
\\
c_1 ={}& 
h_-(l-Q_{N-1})\,h_+(l-Q_{3})\,
g(l)
\;,
\\
c_{2} ={}& h_+(l-Q_{3})\,
h_+(l-Q_{1})\,
g(l)\,
\;,
\\
c_N ={}& h_-(l-Q_{N-1})\,
h_-(l-Q_{1})\,
g(l)\, 
\;,
\\
\tilde c_+ ={}& h_-(l-Q_{N})\,
(x+\bar x)\,\theta(x + \bar x > 0)\
g_-(l)
\;,
\\
\tilde c_- ={}& h_+(l-Q_{2})\,
[-(x+\bar x)]\,\theta(x + \bar x < 0)\
g_+(l)
\;.
\end{split}
\end{equation}
For the special case $A = N-1$, we define
\begin{equation}
\begin{split}
\label{eq:cjnAism1}
c_j ={}& h_-(l-Q_{j-1})\,h_+(l-Q_{j+1})\,
h_-(l-Q_{N})
\\&\times
h_+(l-Q_{N})\,
g(l)\,
\quad j \in \{2,\dots,N-2\}\;,
\\
c_1 ={}& 
h_+(l-Q_{2})\,
h_+(l-Q_{N})\,
g(l)
\;,
\\
c_{N-1} ={}& h_-(l-Q_{N-2})\,
h_-(l-Q_{N})\,
g(l)\, 
\;,
\\
c_N ={}& 
h_+(l-Q_{2})\,h_-(l-Q_{N-2})\,
g(l)\, 
\;,
\\
\tilde c_+ ={}& h_-(l-Q_{N-1})\,
(x+\bar x)\,\theta(x + \bar x > 0)\
g_-(l)
\;,
\\
\tilde c_- ={}& h_+(l-Q_{1})\,
[-(x+\bar x)]\,\theta(x + \bar x < 0)\
g_+(l)
\;.
\end{split}
\end{equation}

The various factors $c_j$ contain factors $h_\pm(l-Q_{i})$ where 
\begin{equation}
\label{eq:hminusdef}
h_-(k) = \frac{(|\vec k| +  E_k)^2 }
{(|\vec k| +  E_k)^2 + M_1^2}\
\theta\!\left(E_k > - |\vec k| \right)
\;\;,
\end{equation}
and
\begin{equation}
\label{eq:hplusdef}
h_+(k) = 
\frac{(|\vec k| -  E_k)^2 }
{(|\vec k| -  E_k)^2 + M_1^2}\
\theta\!\left(E_k < |\vec k| \right)
\;\;.
\end{equation}
Here $E_k$ and $\vec k$ are the energy and three-vector parts of $k$ in a frame in which $P + \bar P$ points along the time axis. Thus $h_-(k) = 0$ for $k \in \bar C_-(0)$ and $h_+(k) = 0$ for $k \in \bar C_+(0)$. These functions depend on the parameter $M_1$ with default value $M_1 = 0.05\,[P\!\cdot\!\bar P]^{1/2}$.

We include in $c_j$ a factor $g(l)$,
\begin{equation}
g(l) = \frac{\gamma_1\,M_2^2}{(l^0-v^0)^2 + (\vec l - \vec v)^2 + M_2^2}
\;\;.
\end{equation}
Here $v$ is the vector defined in Eq.~(\ref{eq:vdpsdef}), $\gamma_1$ is a dimensionless parameter with default value $\gamma_1 = 0.7$, and $M_2$ is a parameter with dimension of mass with default value $M_2 = [P\!\cdot\!\bar P]^{1/2}$.

In $\tilde c_+$ we include a factor $g_-(l)$ and in $\tilde c_-$ we include a factor $g_+(l)$, where
\begin{equation}
g_\pm(l) = 
\frac{\gamma_2}{1 + \left(1 \pm E/\omega\right)^2}
\;\;,
\end{equation}
where $\gamma_2$ is a dimensionless parameter with default value $\gamma_2 = 1$ and
\begin{equation}
\begin{split}
\label{eq:Eomegadef}
E ={}& l^0 - v^0
\;\;,
\\
\omega ={}& \left[(\vec l - \vec v)^2 + M_3^2\right]^{1/2}
\;\;.
\end{split}
\end{equation}
Here $M_3$ is a parameter with dimension of mass with default value $M_3 = [P\!\cdot\!\bar P]^{1/2}$.

Now let us examine what this formula does. We consider first the generic case, $1 < A < N-1$, for which Eq.~(\ref{eq:cjgeneral}) applies. At the end of this section, we discuss the cases $A = 1$ and $A = N-1$.

\subsection{Cone notation}
\label{sec:conenotation}
In order to make the subsequent analysis more compact, we adopt a notation for cones. We denote the forward light cone from $Q_i$,
\begin{equation}
(l-Q_i)^2 = 0,\hskip 1 cm (l-Q_i )\cdot (P+\bar P) > 0
\;\;,
\end{equation}
by $C_+(Q_i)$. Let us also denote by $\bar C_+(Q_i)$ the cone $C_+(Q_i)$ together with its interior:
\begin{equation}
(l-Q_i)^2 \ge 0,\hskip 1 cm (l-Q_i )\cdot (P+\bar P) > 0
\;\;.
\end{equation}
Similarly, we denote the backward light cone from $Q_i$,
\begin{equation}
(l-Q_i)^2 = 0,\hskip 1 cm (l-Q_i )\cdot (P+\bar P) < 0
\;\;,
\end{equation}
by $C_-(Q_i)$ and this cone together with its interior by $\bar C_-(Q_i)$,
\begin{equation}
(l-Q_i)^2 \ge 0,\hskip 1 cm (l-Q_i )\cdot (P+\bar P) < 0
\;\;.
\end{equation}

\subsection{The coefficients $c_j$ for $j \in \{2,\dots,A-1\}$}
\label{sec:leftcase}

Let us examine $c_j$ for $j \in \{2,\dots,A-1\}$. The term in $\kappa_0$ proportional to $c_j$ is
\begin{equation}
\kappa_j = - c_j\ (l - Q_j)
\;\;.
\end{equation}
We need
\begin{equation}
\kappa_j \cdot (l-Q_i) \ge 0
\end{equation}
for $l \in C_\pm(Q_i)$ for any $i$ other than $j$. We have
\begin{equation}
\kappa_{j} \cdot (l - Q_i)
= 
-  c_j (l - Q_i)^2
+  c_j (Q_j - Q_i)\cdot(l - Q_i)
\;\;.
\end{equation}
For $l \in C_\pm(Q_i)$, this becomes
\begin{equation}
\kappa_{j} \cdot (l - Q_i)
= 
c_j (Q_j - Q_i)\cdot(l - Q_i)
\;\;.
\end{equation}

Consider first the case $j < i \le A$. Then $Q_j \in \bar C_-(Q_i)$. For $l \in C_-(Q_i)$, we then have $\kappa_{j} \cdot (l - Q_i) \ge 0$, as required (since we define $c_j$ so that it is non-negative).  For $l \in C_+(Q_i)$, we have the ``wrong'' sign, $\kappa_{j} \cdot (l - Q_i) \le 0$, so we need to define $c_j$ so that it vanishes for $l \in C_+(Q_i)$. We note that $C_+(Q_i) \subset \bar C_+(Q_{j+1})$. Thus we ensure that $c_j$ vanishes for $l \in C_+(Q_i)$ by including a factor $h_+(l-Q_{j+1})$ in $c_j$.

Consider next the case $1 \le i < j$. Then $Q_j \in \bar C_+(Q_i)$. For $l \in C_+(Q_i)$, we then have $\kappa_{j} \cdot (l - Q_i) \ge 0$, as required.  For $l \in C_-(Q_i)$, we have the ``wrong'' sign, $\kappa_{j} \cdot (l - Q_i) \le 0$, so we need to define $c_j$ so that it vanishes for $l \in C_-(Q_i)$. We note that $C_-(Q_i) \subset \bar C_-(Q_{j-1})$. Thus we ensure that $c_j$ vanishes for $l \in C_-(Q_i)$ by including a factor $h_-(l-Q_{j-1})$ in $c_j$.

Finally, consider the case $A+1 \le i \le N$. In this case, $Q_j - Q_i$ is spacelike, so that $\kappa_{j} \cdot (l - Q_i)$ can have either sign for $l \in C_\pm(Q_i)$. Thus we need to define $c_j$ so that it vanishes for $l \in C_\pm(Q_i)$. We note that $C_+(Q_i) \subset \bar C_+(Q_{A+1})$ and $C_-(Q_i) \subset \bar C_+(Q_{N})$. Thus we ensure that $c_j$ vanishes for $l \in C_\pm(Q_i)$ by including a factors $h_+(l-Q_{A+1})$ and $h_-(l-Q_{N})$ in $c_j$.

We also include a factor $g(l)$ in $c_j$. The purpose of the factor $g(l)$ is to turn off the deformation when $\vec l$ and $l^0$ are large.

\subsection{The coefficient $c_j$ for $j = A$.}
\label{sec:leftcaseA}

We have discussed $c_j$ for $j \in \{2,\dots,A-1\}$. The case $j = A$ is similar, but not identical.

There are no cones $C_\pm(Q_i)$ for $j < i \le A$. Therefore we do not need a factor  $h_+(l-Q_{j+1})$ in $c_j$. We simply omit this factor. We do include a factor $h_-(l-Q_{j-1})$, as before.

For $A+2 \le i \le N$, $Q_j - Q_i$ is spacelike, so that $\kappa_{j} \cdot (l - Q_i)$ can have either sign for $l \in C_\pm(Q_i)$. Thus we need to define $c_j$ so that it vanishes for $l \in C_\pm(Q_i)$. We note that $C_+(Q_i) \subset \bar C_+(Q_{A+2})$ and $C_-(Q_i) \subset \bar C_+(Q_{N})$. Thus we ensure that $c_j$ vanishes for $l \in C_\pm(Q_i)$ by including a factors $h_+(l-Q_{A+2})$ and $h_-(l-Q_{N})$ in $c_j$.

For $i = A+1$, $Q_j - Q_i$ is lightlike, $Q_j \in C_+(Q_i)$. For $l \in C_+(Q_i)$, we then have $\kappa_{j} \cdot (l - Q_i) \ge 0$, as required.  For $l \in C_-(Q_i)$, we have the ``wrong'' sign, $\kappa_{j} \cdot (l - Q_i) \le 0$, so we need to define $c_j$ so that it vanishes for $l \in C_-(Q_i)$. We note that $C_-(Q_i) \subset \bar C_-(Q_{N})$. Thus the factor $h_-(l-Q_{N})$ in $c_j$ ensures $c_j$ vanishes for $l \in C_-(Q_i)$. 

We include a factor $g(l)$ in $c_j$ as before.

\subsection{The coefficient $c_j$ for $j = 1$.}
\label{sec:leftcase1}

The case $j = 1$ is essentially the same as the case $j=A$. We do not need a factor $h_-(l-Q_{j-1})$ in $c_j$ because there are no cones $C(Q_i)$ for $1 \le i < j$. We ensure that $c_j = 0$ for $l \in C_\pm(Q_i)$ for $A + 1 \le i \le N-1$ and for $l \in C_+(Q_N)$ by including factors  $h_+(l-Q_{A+1})$ and $h_-(l-Q_{N-1})$ in $c_j$.

\subsection{The coefficients $c_j$ for $j \in \{A+1,\dots,N\}$}
\label{sec:rightcase}

We define the coefficients $c_j$ for $j \in \{A+1,\dots,N\}$ according to the same pattern that we used for $j \in \{1,\dots,A\}$.

\subsection{The coefficient $\tilde c_+$}
\label{sec:topcase}

The deformations $\kappa_j = - c_j\,(l - Q_j)$ have been arranged so that $(l-Q_i)\cdot \kappa_j \ge 0$ for $l \in C_\pm(Q_i)$. We in fact want $(l-Q_i)\cdot \sum \kappa_j > 0$ except on the lightlike lines along which the contour is pinched. A straightforward but tedious analysis shows that we have achieved that objective in the left region and in the right region in Fig.~\ref{fig:Qnplus}. However, we have $(l-Q_i)\cdot \sum \kappa_j = 0$ on the cones $C_\pm(Q_i)$ in a large part of the top region. This is because $\sum \kappa_j = 0$ in the intersection of $C_+(Q_2)$ and $C_+(Q_{A+2})$.

This deficiency is easy to fix. In the top region, on one of the cones $C_+(Q_i)$ we have $\kappa\cdot (l-Q_i) > 0$ for any timelike $\kappa$ with $\kappa^0 > 0$. In particular, one could take $\kappa \propto P+\bar P$. We therefore add a term
\begin{equation}
\tilde c_+\,(P + \bar P)
\end{equation}
to $\kappa_0$. We include in the coefficient $\tilde c_+$ factors $h_-(l - Q_A)$ and $h_-(l - Q_N)$ so that $\tilde c_+ = 0$ on all of the backward light cones $C_-(Q_i)$, for which $P + \bar P$ is in the ``wrong'' direction.

We also include in $\tilde c_+$ factors $(x+\bar x)\,\theta(x + \bar x > 0)$ and  $g_-(l)$. The purpose of these factors is to control the deformation in the region of large momenta. Near the forward light cones $C_+(Q_i)$ we have $E \sim \omega$. In this region, $g_-(l) \sim 1$, while $(x+\bar x)$ grows with $l$. This gives us a big deformation that can keep the contour well away from these light cones. However, for any fixed $\vec l$, $(x+\bar x)\,g_-(l) \to 0$ as $E \to \infty$, thus turning the deformation off.

\subsection{The coefficient $\tilde c_-$}
\label{sec:bottomcase}

In the bottom region, on one of the cones $C_-(Q_i)$ we have $\kappa\cdot (l-Q_i) > 0$ for any timelike $\kappa$ with $\kappa^0 < 0$. In particular, one could take $\kappa \propto -(P+\bar P)$. We therefore add a term
\begin{equation}
-\tilde c_-\,(P + \bar P)
\end{equation}
to $\kappa_0$. We include in the coefficient $\tilde c_-$ factors $h_+(l - Q_1)$ and $h_+(l - Q_{A+1})$ so that $\tilde c_- = 0$ on all of the forward light cones $C_+(Q_i)$, for which $-(P + \bar P)$ is in the ``wrong'' direction.

We also include in $\tilde c_-$ factors $-(x+\bar x)\,\theta(x + \bar x < 0)$ and  $g_+(l)$ that serve to control the deformation in the region of large momenta.

\subsection{The special case $A = 1$}
\label{sec:Ais1}

The preceding discussion has concerned the generic case $1 < A < N-1$. When $A = 1$, the situation is a little different.

The definition of $c_j$ for $j \in \{3,\dots,N-1\}$ follows the logic of the generic case.

For $c_1$, we note that $\kappa_1 \propto -(l - Q_1)$ points in the right direction on $C_-(Q_N)$ and on $C_+(Q_2)$ but in the wrong direction on $C_-(Q_i)$ for $2 \le i \le N-1$ and on $C_+(Q_i)$ for $3 \le i \le N$. We arrange that $c_1$ vanishes on the cones $C_-(Q_i)$ for $2 \le i \le N-1$ and on $C_+(Q_i)$ for $3 \le i \le N$ by including factors $h_-(l-Q_{N-1})$ and $h_+(l-Q_3)$ in $c_1$.

For $c_2$, we note that $\kappa_2 \propto -(l - Q_2)$ points in the right direction on $C_-(Q_1)$ and on $C_-(Q_i)$ for $3 \le i \le N$. However, $\kappa_2$ points in the wrong direction on $C_+(Q_1)$ and on $C_+(Q_i)$ for $3 \le i \le N$. We arrange that $c_2$ vanishes on the cones $C_+(Q_1)$ and on $C_+(Q_i)$ for $3 \le i \le N$ by including factors $h_+(l-Q_{1})$ and $h_+(l-Q_3)$ in $c_2$.

For $c_N$, the same logic leads us to include factors $h_-(l-Q_{1})$ and $h_-(l-Q_{N-1})$ in $c_N$.

For $\tilde c_+$, we note that $P+\bar P$ points in the right direction on all of the cones $C_+(Q_i)$ but in the wrong direction on all of the cones $C_-(Q_j)$. We arrange that $\tilde c_+$ vanish on all of the cones $C_-(Q_j)$ by including a factor $h_-(l-Q_N)$ in $\tilde c_+$. We also include in $\tilde c_+$ factors $(x+\bar x)\,\theta(x + \bar x > 0)$ and  $g_-(l)$, as before.

The construction of $\tilde c_-$ folllows the logic used for $\tilde c_+$.

\subsection{The special case $A = N-1$}
\label{sec:AisN}

The construction of the deformation when $A = N-1$ follows the logic of the case $A = 1$.

\section{How far to deform}

We now take the deformation to be $l \to l + i \kappa$, where 
\begin{equation}
\kappa = \lambda \kappa_0
\end{equation}
and $\lambda$ is a scalar normalization factor. We must ensure that the integrand does {\em not} have any singularities as $\lambda$ varies from zero to its chosen positive real value. 

It is not immediately obvious how to achieve this end. If we take a very small value of $\lambda$ then we have effectively an infinitesimal deformation. We have arranged that our deformation for small $\lambda$ is in the right direction, so we can be sure that we have avoided all singularities except those that cannot be avoided because they are pinch singularities. On the other hand, we come very close to the singularities, which is not good for the numerical convergence of the integration. Thus we need to make $\lambda$ as large as we can. Then the integration contour is far from the product of the real $l^\mu$ axes and it is not so easy to see how to guarantee that in deforming away from one cone $(l + i \lambda \kappa_0 - Q_i)^2 = 0$ we do not run into another cone $(l + i \lambda \kappa_0 - Q_j)^2 = 0$.

To ensure that the integrand does {\em not} have any singularities as $\lambda$ varies from zero to its chosen positive real value, we can take $\lambda$ to be the smallest of a number, $\lambda_i(l)$, defined for each propagator, a general choice, $\lambda_0(l)$, to be fixed later, and a fixed, $l$ independent, constant $\lambda_{\rm c}$ (the default is $\lambda_{\rm c} = 1$):
\begin{equation}
\lambda(l) = \min[\lambda_{\rm c},\lambda_0(l),\min_i\{\lambda_i(l)\}]\;\;.
\end{equation}
In order to examine what $\lambda_i$ ought to be, we consider the $i$th denominator,
\begin{equation}
D_i = (l - Q_i + \mi \lambda\kappa_0)^2
= (l - Q_i)^2 + 2 \mi \lambda\kappa_0\cdot(l - Q_i)
- \lambda^2 \kappa_0^2
\;\;.
\end{equation}
We note that the function $D_i$ vanishes at values of $\lambda$ given by
\begin{equation}
\begin{split}
\lambda ={}& \frac{1}{\kappa_0^2}\ \bigg\{
\mi\ \kappa_0\cdot(l - Q_i)
\\&
\pm\sqrt{\kappa_0^2\,(l - Q_i)^2 - [\kappa_0\cdot(l-Q_i)]^2}
\bigg\}
\;\;.
\end{split}
\end{equation}

Consider what happens if $\kappa_0^2\,(l - Q_i)^2 > [\kappa_0\cdot(l-Q_i)]^2$. Then one of the poles crosses the real $\lambda$ axis when, as we vary $l$, $\kappa_0\cdot(l - Q_i)$ crosses zero. The absolute value of the pole position is
\begin{equation}
\begin{split}
|\lambda|& =
\\&
\frac{1}{|\kappa_0^2|}
\sqrt{[\kappa_0\cdot(l-Q_i)]^2 
+ \left[\kappa_0^2\,(l - Q_i)^2 - (\kappa_0\cdot(l-Q_i))^2
\right]}
\\&
=
\sqrt{\frac{(l - Q_i)^2}{\kappa_0^2}}
\;\;. 
\end{split}
\end{equation}

There is a dangerous region $[\kappa_0\cdot(l-Q_i)]^2 \ll \kappa_0^2\,(l - Q_i)^2$. In this dangerous region, one of the two poles is near the positive real $\lambda$ axis. If we increase $\lambda$ from 0 past the real part of the pole position, we come very near the pole. If $\kappa_0\cdot(l-Q_i) = 0$, we pass {\em through} the pole. Thus we need to keep the actual value of $\lambda$ smaller than the real part of the pole position when the parameters $(\kappa_0\cdot(l-Q_i), \kappa_0^2\,(l - Q_i)^2)$ are in the dangerous region.

Consider also the case $\kappa_0^2\,(l - Q_i)^2 < [\kappa_0\cdot(l-Q_i)]^2$. Then $D_i$ vanishes at values of $\lambda$ given by
\begin{equation}
\begin{split}
\label{eq:lambdasoln}
\lambda ={}& \frac{\mi}{\kappa_0^2}\ \bigg\{
\kappa_0\cdot(l - Q_i)
\\&\quad
\pm\sqrt{[\kappa_0\cdot(l-Q_i)]^2 - \kappa_0^2\,(l - Q_i)^2}
\bigg\}
\;\;.
\end{split}
\end{equation}

Both poles are on the imaginary $\lambda$ axis and there is no limitation on how big a positive real value value one can take for $\lambda$. In fact, choosing a bigger value for $\lambda$ moves us further away from the poles.

We now define $\lambda_i$. We first define $\lambda_i$ for the region $2 [\kappa_0\cdot(l-Q_i)]^2 < \kappa_0^2\,(l - Q_i)^2$, which includes the dangerous region $[\kappa_0\cdot(l-Q_i)]^2 \ll \kappa_0^2\,(l - Q_i)^2$. In this region, to make sure that $\lambda_i$ is small enough, we let it be half the absolute value of the pole position. That is
\begin{equation}
\begin{split}
\label{eq:lambdadef1}
\lambda_i^2 ={}& 
\frac{\kappa_0^2(l - Q_i)^2}{(2\kappa_0^2)^2}
\\&
\quad {\rm for}\ 2[\kappa_0\cdot(l-Q_i)]^2 < \kappa_0^2\,(l - Q_i)^2
\;\;.
\end{split}
\end{equation}
For the region $0 < \kappa_0^2\,(l - Q_i)^2 < 2[\kappa_0\cdot(l-Q_i)]^2$, we define
\begin{equation}
\begin{split}
\label{eq:lambdadef2}
\lambda_i^2 ={}& 
\frac{4 [\kappa_0\cdot(l-Q_i)]^2 - \kappa_0^2(l - Q_i)^2}{(2\kappa_0^2)^2}
\\&
\quad {\rm for}\ 0 < \kappa_0^2\,(l - Q_i)^2 < 2[\kappa_0\cdot(l-Q_i)]^2
\;\;.
\end{split}
\end{equation}
Note that the definitions (\ref{eq:lambdadef1}) and (\ref{eq:lambdadef2}) match at $\kappa_0^2\,(l - Q_i)^2 = 2[\kappa_0\cdot(l-Q_i)]^2$. For the region $\kappa_0^2\,(l - Q_i)^2 < 0$, we define
\begin{equation}
\begin{split}
\label{eq:lambdadef3}
\lambda_i^2 ={}& 
\frac{4 [\kappa_0\cdot(l-Q_i)]^2 - 2\kappa_0^2(l - Q_i)^2}{(2\kappa_0^2)^2}
\\&
\quad {\rm for}\ \kappa_0^2\,(l - Q_i)^2 < 0
\;\;.
\end{split}
\end{equation}
Note that the definitions (\ref{eq:lambdadef2}) and (\ref{eq:lambdadef3}) match at $\kappa_0^2\,(l - Q_i)^2 = 0$.

This method needs some special consideration for the case of $\lambda_i$ when $l$ approaches one of the collinear singularities along the two lines $l = Q_i + z (Q_{i+1}-Q_i)$ and $l = Q_i + z (Q_{i-1}-Q_i)$ that meet at $l = Q_i$. In these cases, the numerator in the definition of $\lambda_i$ vanishes in the limit and also the denominator, $\kappa_0^2$, vanishes in the limit. Potentially, then, $\lambda_i$ approaches zero in the limit and its derivatives with respect to $l$ are singular. This could lead to a singularity in the jacobian for the deformation, the determinant of the matrix $\delta^\mu_\nu + i \partial \kappa^\mu/\partial l^\nu$.

To analyze this, we write
\begin{equation}
\begin{split}
\kappa_0 ={}& -\sum_{j=1}^N c_j (l-Q_j)
+ \tilde c_+ (P + \bar P)
- \tilde c_- (P + \bar P)
\\
={}& - C\, (l-Q_i) + R_i
\;\;,
\end{split}
\end{equation}
where
\begin{equation}
\begin{split}
C ={}& \sum_{j=1}^N c_j
\;\;,
\\
R_i ={}& \sum_{j=1}^N c_j (Q_j - Q_i)
+ \tilde c_+ (P + \bar P)
- \tilde c_- (P + \bar P)
\;\;.
\end{split}
\end{equation}
We note that
\begin{equation}
\kappa_0\cdot(l - Q_i) = -C\,(l-Q_i)^2
+ (l-Q_i)\cdot R_i
\;\;,
\end{equation}
while
\begin{equation}
\kappa_0^2 = C^2\,(l-Q_i)^2
- 2 C\, (l-Q_i)\cdot R_i
+ R_i^2
\;\;.
\end{equation}
We can eliminate $(l-Q_i)\cdot R_i$ between these to obtain
\begin{equation}
(l-Q_i)^2 =
-\frac{1}{C^2}\,\kappa_0^2 
- \frac{2}{C}\,\kappa_0\cdot (l-Q_i) 
+\frac{1}{C^2}\,R_i^2
\;\;.
\end{equation}
So far, this is general. If we now consider the approach to the collinear singularity, we have $R_i^2 \to 0$. Furthermore, $R_i^2$ approaches zero fast enough, like $[(l-Q_i)^2]^2$, that we can set it to zero here. Then
\begin{equation}
\label{eq:kinematicsline}
2\kappa_0\cdot (l-Q_i)
 \approx
- C(l-Q_i)^2
-\frac{1}{C}\,\kappa_0^2  
\;\;.
\end{equation}
Using this relation, we can see that when $l$ is close to the collinear singularity, it is never in the region $2 [\kappa_0\cdot(l-Q_i)]^2 < \kappa_0^2\,(l - Q_i)^2$ for which the choice (\ref{eq:lambdadef1}) for $\lambda$ applies. Indeed, squaring Eq.~(\ref{eq:kinematicsline}) gives
\begin{equation}
\begin{split}
\label{eq:kinematicsline2}
2[\kappa_0\cdot (l-Q_i)]^2
 \approx{}&
\kappa_0^2(l-Q_i)^2
\\&
+\frac{C^2}{2}[(l-Q_i)^2]^2
+\frac{1}{2C^2}\left[\kappa_0^2 \right]^2
\\
>{}& \kappa_0^2(l-Q_i)^2
\;\;.
\end{split}
\end{equation}

When $l$ is near the collinear singularity, it can be in either of the regions for which the choices for $\lambda_i$ given in Eqs.~(\ref{eq:lambdadef2}) and (\ref{eq:lambdadef3}) apply, depending on the sign of $\kappa_0^2\,(l - Q_i)^2$. If we use Eq.~(\ref{eq:kinematicsline}) in Eq.~(\ref{eq:lambdadef2}), we have, for $\kappa_0^2\,(l - Q_i)^2 > 0$,
\begin{equation}
\begin{split}
\label{eq:lambdadef2bis}
\lambda_i^2 ={}& 
\frac{
\kappa_0^2(l-Q_i)^2
+ C^2[(l-Q_i)^2]^2
+ \left[\kappa_0^2 \right]^2 /C^2
}{
4 (\kappa_0^2)^2
}
\\
>{}& \frac{1}{4 C^2}
\;\;. 
\end{split}
\end{equation}
If we use Eq.~(\ref{eq:kinematicsline}) in Eq.~(\ref{eq:lambdadef3}), we have, for $\kappa_0^2\,(l - Q_i)^2 < 0$,
\begin{equation}
\label{eq:lambdadef3bis}
\lambda_i^2 = 
\frac{
C^2[(l-Q_i)^2]^2
+ \left[\kappa_0^2 \right]^2 /C^2
}
{4(\kappa_0^2)^2}
> \frac{1}{4 C^2}
\;\;.
\end{equation}
Thus near the either of the collinear singularities along lines that meet at $l = Q_i$ we have
\begin{equation}
\lambda_i > \frac{1}{2C}
\;\;.
\end{equation}

We want to ensure that $\lambda$ is continuous as one approaches the collinear singularity. We define
\begin{equation}
\lambda_0(l) = \frac{1}{4\,C(l)}
\;\;.
\end{equation}
Then we will have $\lambda = \lambda_0 = {1}/(4\,C)$, which is a smooth function of $l$, near the collinear singularity.

\section{Results}

We have constructed computer code \cite{whereiscode} that carries out the calculations outlined here. In this section, we use this code to test how well the method described works. We calculate the sum over graphs of ${\cal M}$ as specified in Eq.~(\ref{eq:calM0}). For $N$ photons, there are $(N-1)!$ graphs obtained by taking non-cycic permutations of the momenta $p_i$ and photon polarization vectors $\epsilon_i$.\footnote{With vector electromagnetic currents, charge conjugation invariance implies that graphs that differ by reversing the order of the external photons are equal. Thus one might say that there are $(N-1)!/2$ independent graphs.} We start with a labeling in which photons 1 and 2 are the incoming particles and photons $3,\dots,N$ are outgoing. Then we sum over graphs by summing over the results obtained with non-cyclic permutations of the indices $i$.

The result for a given choice of helicities of the photons has a phase that depends on the precise definition of the photon polarization vectors $\epsilon_i$. However, the absolute value of the scattering amplitude ${\cal M}$ is independent of the conventions used to define the $\epsilon_i$, so we concentrate on $|{\cal M}|$.  Since $|{\cal M}|$ is proportional to $\alpha^{N/2}$ and has mass dimension $4-N$, we exhibit $|{\cal M}|\times (\sqrt{s})^{N-4}/\alpha^{N/2}$ in our plots. We specify helicities in the form $h_1,h_2,h_3,\dots,h_N$, where 1 and 2 are the incoming particles and, following convention, $h_1$ and $h_2$ are actually the negative of the physical helicities of the incoming photons.

We compute ${\cal M}$ by Monte Carlo numerical integration on the deformed contour as described in this paper. The integration samples points $l$ giving extra weight to regions near the intersections of two or three light cones and to the region near the double parton scattering configuration. Of the $(N-1)!$ graphs, most weight is given to graphs for which the corresponding integral is closest to being singular. The code at \cite{whereiscode} comes with some documentation of the sampling method.

\subsection{$N = 6$}
\label{sec:results6}

We compute the six photon amplitude along a certain one dimensional curve in the space of final state momenta. We take photon 1 to have momentum $\vec p_1$ along the $-z$-axis (so the physical incoming momentum is along the $+z$-axis), and we take $\vec p_2$ along the $+z$-axis. We choose an arbitrary point for the final state momenta $\{\vec p_3, \vec p_4, \vec p_5, \vec p_6\}$:
\begin{equation}
\begin{split}
\vec p_3 ={}& (33.5,15.9,25.0)
\;\;,
\\
\vec p_4 ={}& (-12.5,15.3,0.3)
\;\;,
\\
\vec p_5 ={}& (-10.0,-18.0,-3.3)
\;\;,
\\
\vec p_6 ={}& (-11.0,-13.2,-22.0)
\;\;.
\end{split}
\end{equation}
Then we create new momentum configurations by rotating the final state through angle $\theta$ about the $y$-axis.

\begin{figure}
\centerline{
\includegraphics[width = 8 cm]{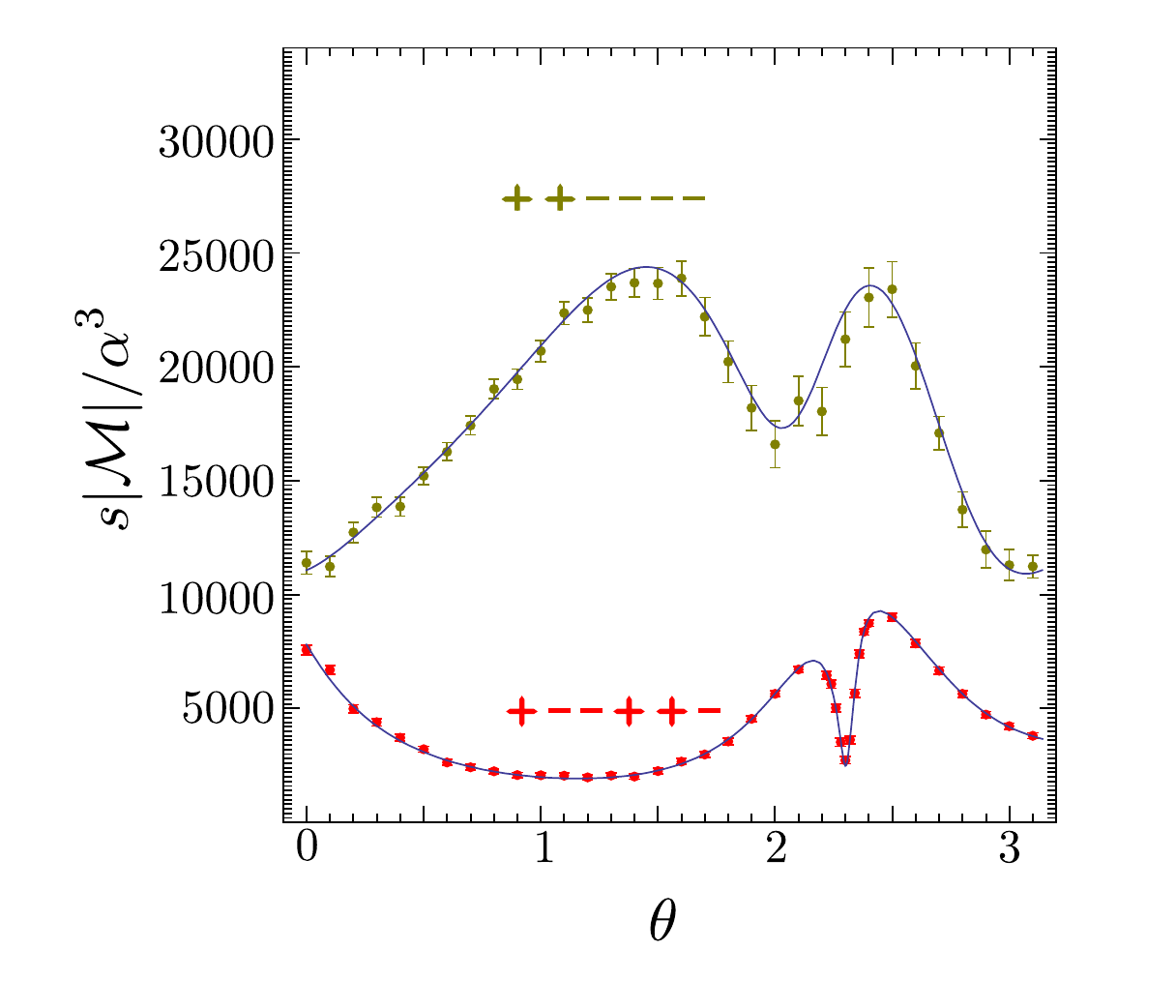}
}
\medskip
\caption{
Results for the six photon amplitude. An arbitrarily chosen final state was rotated about the $y$-axis through angle $\theta$. The numerical results for the helicity choice $+$$+$$-$$-$$-$$-$ are compared with the analytic result of Ref.~\cite{Mahlon}. The numerical results for the helicity choice $+$$-$$-$$+$$+$$-$ are compared with the analytic result of Ref.~\cite{Binoth}. The numerical results were generated using $10^6$ Monte Carlo points for each of 120 graphs.
}
\label{fig:sixphotonsNEW}
\end{figure}

We compute the six photon amplitude by Monte Carlo numerical integration on the deformed contour as described in this paper. In Fig.~\ref{fig:sixphotonsNEW}, we plot computed values of $s\,|{\cal M}|/\alpha^{3}$ versus $\theta$ in the range from $0$ to $\pi$. We show numerical results for the helicity choices $+$$+$$-$$-$$-$$-$ and $+$$-$$-$$+$$+$$-$. For the helicity choice $+$$+$$-$$-$$-$$-$, we compare the results with the analytic result of Mahlon, \cite{Mahlon}. For the helicity choice $+$$-$$-$$+$$+$$-$, we compare the results with the recent analytic result of Binoth, Heinrich, Gehrmann and Mastrolia \cite{Binoth}, who also confirm the result for $+$$+$$-$$-$$-$$-$. We see that the numerical results agrees with the analytical results. At $\theta \approx 2.32$, the external momentum configuration lies close to a double parton scattering singularity: $(p_{{\rm T},3} + p_{{\rm T},5})^2 \approx 0.0003\,s$. We note that there is a quite sharp structure in $|{\cal M}|$ near this angle for the helicity choice $+$$-$$-$$+$$+$$-$. The numerical results were generated using $10^6$ Monte Carlo points for each of 120 graphs.

Ref.~\cite{Mahlon} shows analytically that $\cal M$ vanishes for the helicity choices $+$$+$$+$$+$$+$$+$ and $+$$+$$+$$+$$+$$-$. We confirm that the numerical integration gives zero within errors for these helicity choices.

\begin{figure}
\centerline{
\includegraphics[width = 8 cm]{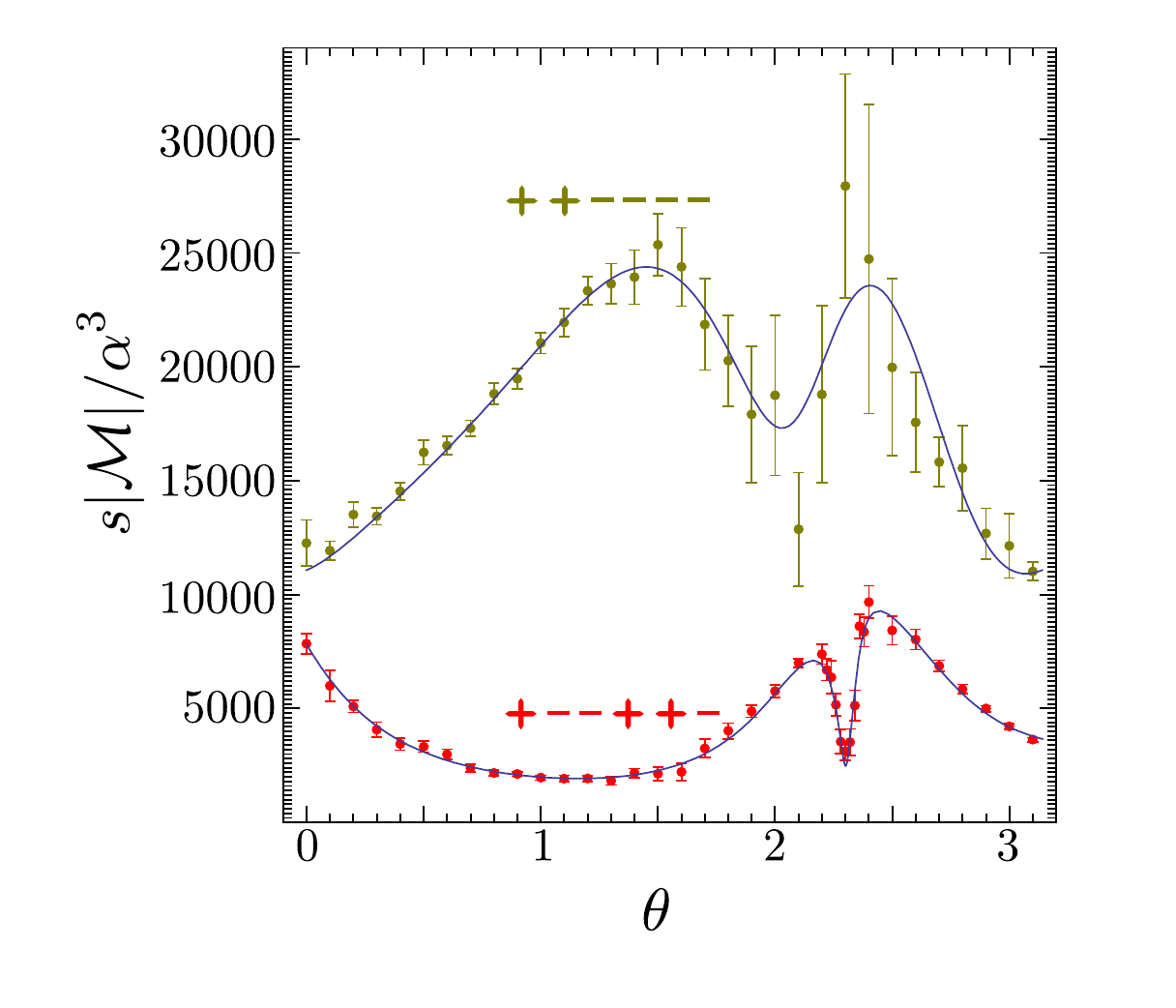}
}
\medskip
\caption{
Results for the six photon amplitude using the Feynman parameter representation, Eq.~(\ref{eq:lxspacedeformed}). The labeling is as in Fig.~\ref{fig:sixphotonsNEW}.
}
\label{fig:sixphotonsOLD}
\end{figure}

Compared to the method of Ref.~\cite{NSnumerical} that applies Monte Carlo numerical integration to the Feynman parameter representation of Eq.~(\ref{eq:lxspacedeformed}), how practical is the method presented in this paper? To see, we present in Fig.~\ref{fig:sixphotonsOLD} numerical results using the Feynman parameter representation. The numerical results for $0 \le \theta \le 2.0$ were published in Ref.~\cite{NSnumerical}. For each of these angles, $1\times 10^6$ Monte Carlo points were used for each of 120 graphs. The points for $2.0 < \theta  < \pi$ were added later after the analytical results of Ref.~\cite{Binoth} were published. For these angles, we used $3\times 10^6$ Monte Carlo points for each graph in order to explore with higher accuracy the region near $\theta = 2.3$, which is close to a double parton scattering singularity. The running time per Monte Carlo point in the Feynman parameter program is roughly half that of the direct momentum space method. Thus running times in Fig.~\ref{fig:sixphotonsNEW} are comparable to those for Fig.~\ref{fig:sixphotonsOLD}. Comparing the error bars, we conclude that the direct momentum space method somewhat outperforms the Feynman parameter method.

\subsection{$N=8$}

We have also computed the eight photon amplitude along a one dimensional curve in the space of final state momenta, as for six photons. We take photon 1 to have momentum $\vec p_1$ along the $-z$-axis and we take $\vec p_2$ along the $+z$-axis. We choose an arbitrary point for the final state momenta $\{\vec p_3, \vec p_4, \vec p_5, \vec p_6, \vec p_7, \vec p_8\}$:
\begin{equation}
\begin{split}
\vec p_3 ={}& (33.5,5.9,25.0)
\;\;,
\\
\vec p_4 ={}& (1.5,24.3,0.3)
\;\;,
\\
\vec p_5 ={}& (-19.1,-35.1,-3.3)
\;\;,
\\
\vec p_6 ={}& (28.2,-6.6,8.2)
\;\;,
\\
\vec p_7 ={}& (-12.2,-8.6,8.2)
\;\;,
\\
\vec p_8 ={}& (-31.9,20.1,-38.4)
\;\;.
\end{split}
\end{equation}
Then we create new momentum configurations by rotating the final state through angle $\theta$ about the $y$-axis. 

\begin{figure}
\centerline{
\includegraphics[width = 9 cm]{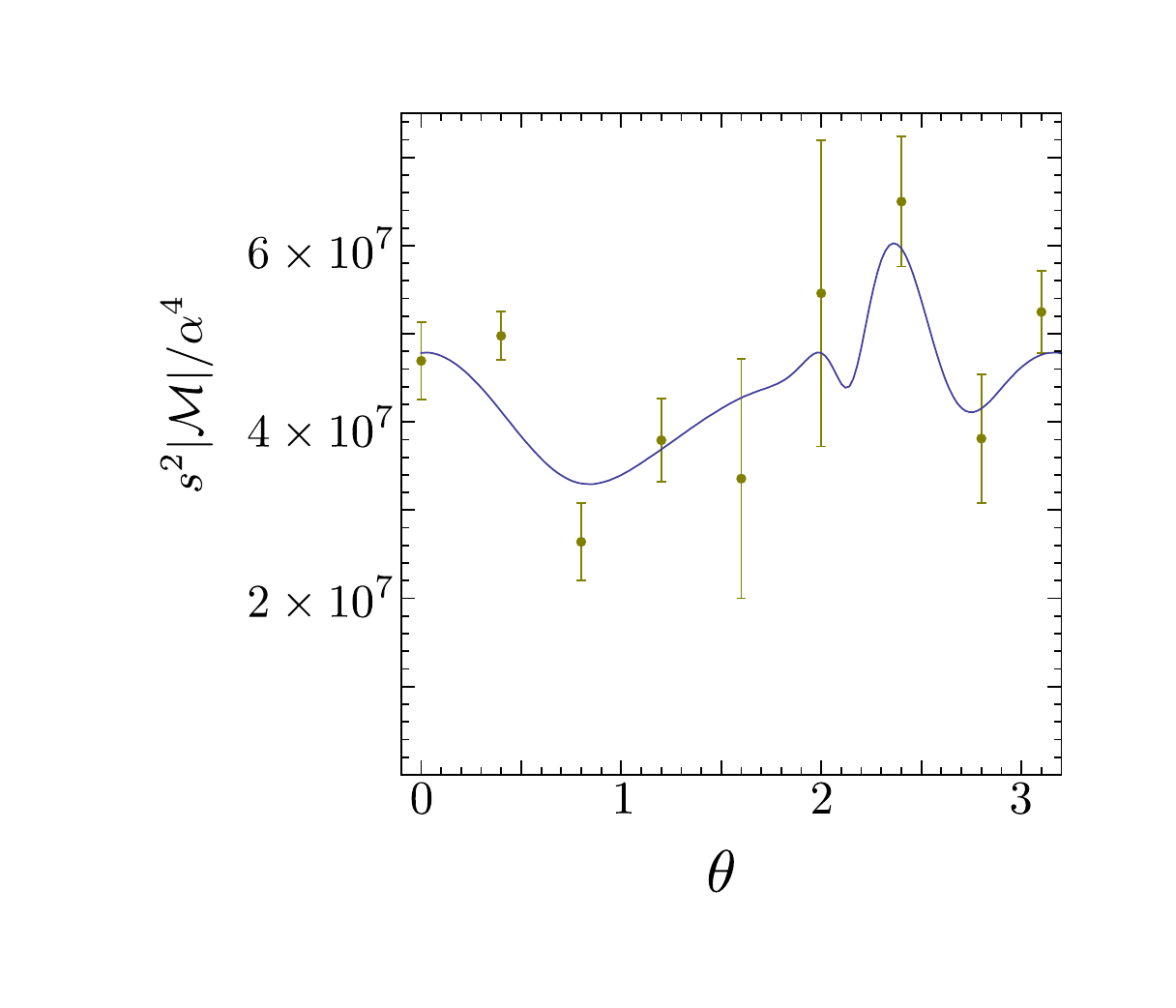}
}
\medskip
\caption{
Results for the eight photon amplitude. An arbitrarily chosen final state was rotated about the $y$-axis through angle $\theta$. The numerical results for the helicity choice $+$$+$$-$$-$$-$$-$$-$$-$ are compared with the analytic result of Ref.~\cite{Mahlon}. For most of the chosen values of $\theta$, the numerical results were generated using $2\times 10^5$ Monte Carlo points for each of 5040 graphs. For $\theta = 2.0$ and $\theta = 2.4$, $10^6$ points were used for each graph.
}
\label{fig:eightphotons}
\end{figure}

We compute the six photon amplitude by Monte Carlo numerical integration on the deformed contour as described in this paper. In Fig.~\ref{fig:eightphotons}, we plot computed values of $s^2\,|{\cal M}|/\alpha^{4}$ versus $\theta$ in the range from $0$ to $\pi$. We show numerical results for the helicity choices $+$$+$$-$$-$$-$$-$$-$$-$ and compare the results with the analytic result of Mahlon \cite{Mahlon}. There are 5040 graphs and we used $2\times 10^5$ Monte Carlo points for each of them except at the angles $\theta = 2.0$ and $\theta = 2.4$, where we used $10^6$ points for each graph. We agree with the analytic results. As one can see from the figure, the numerical convergence of the integration is not as good with eight photons as it was for six. This should not be a surprise. There are now more graphs and, in each graph, there are now eight propagators that can be singular. The contour deformation allows us to escape from singularities, but the more propagators there are, the less effective this escape is. 

Despite the marginal convergence of the integration with eight photons, we can note that the deformation method is quite robust. We have simply given it a hard test problem. The integration method of Ref.~\cite{NSnumerical}, involving numerical integration using Feynman parameters, could not produce any results at all with eight photons.

\section{Conclusions}

We have investigated how to perform the integral
\begin{equation}
{\cal M} = \int\! \frac{d^4 l}{(2\pi)^4}\ (-\mi e)^N N(l)
\prod_{i=1}^N \frac{\mi}{(l - Q_i)^2 + \mi 0}
\label{eq:calM0bis}
\end{equation}
for the $N$-photon scattering amplitude with a massless electron loop using numerical Monte Carlo integration. The main ingredient needed to perform the integration is to deform the integration contour into the space of complex momenta $l + i \kappa$ in such a way as to avoid the singularities of the denominators wherever the contour is not pinched.

We have in mind applications of the method described here to next-to-leading order calculations of infrared-safe observables in the Standard Model and its extensions. For this application, the integral of Eq.~(\ref{eq:calM0bis}) is embedded in a larger integral and the whole integral is performed by numerical Monte Carlo integration. Generally, infrared subtractions are needed. These can be obtained for virtual loops from a construction \cite{NSsubtractions} that is similar to the construction commonly used for infrared singularities in real emission diagrams. The numerical integration method for virtual loop integrals has proved useful, as described in the Introduction, but it is not clear to us that this method is more practical than methods that calculate the amplitude ${\cal M}$ as a whole and insert the complete value of  ${\cal M}$ into the larger integral that finally gives a physical observable. We offer this method because we believe that it is good to have a range of methods available.

The method described here is similar to the method of Ref.~\cite{NSnumerical}, in which the amplitude (\ref{eq:calM0bis}) is first represented as an integral over Feynman parameters. The Feynman parameter method involves simpler contour deformations. On the other hand, our numerical results suggest that the convergence properties of the integral are better with the present ``direct'' method.

We have seen how to deform the integration contour for the problem posed, with massless particles. This is an important case for applications to high energy scattering. It is also a case with some difficulties because with massless particles there are soft and collinear singularities in the integrand. We leave for future work cases with particle masses. With masses, the cones that appear in this paper turn into hyperboloids, requiring a different choice of contour deformations. Assuming that one has redefined the contour deformations, having non-zero masses can make the problem easier since it eliminates some or all of collinear and soft singularities. On the other hand, the general problem with masses is more difficult because threshold singularities can occur. 

We also note that there could well be applications in which one wants to compute integrals like (\ref{eq:calM0bis}) as a whole, without needing to insert the integral into something else. The completely numerical method presented here has an advantage of simplicity: the method does not depend on the numerator function $N(l)$. One might hope to have a general method that constructs the proper contour deformations for the denominators, with masses, as we have done here for the massless case.

\acknowledgements

This work was supported in part by the United States Department of Energy, the Helmoltz Alliance network ``Physics at the Terascale," and the Hungarian Scientific Research Fund grant OTKA K-60432. We thank T.~Binoth and G.~Heinrich for providing results from Ref.~\cite{Binoth} in a convenient form.


\end{document}